\documentclass[a4paper, twocolumn, final]{elsarticle}

\journal{Medical Image Analysis}
\usepackage{makeidx}  
\usepackage{color,xcolor,soul,colortbl}

\usepackage{bm}
\usepackage{hyphenat}
\biboptions{sort,comma,authoryear}

\usepackage{mathtools}
\usepackage{amsmath}
\DeclareMathOperator{\std}{\mathrm{std}}

\usepackage{float}
\usepackage{graphicx,subfigure,psfrag,epsfig}
\graphicspath{{figures/}}
\usepackage{sidecap} 	
\usepackage{epstopdf}
\usepackage[labelfont=bf]{caption}
\usepackage{array}
\newcolumntype{L}[1]{>{\raggedright\let\newline\\\arraybackslash\hspace{0pt}}m{#1}}
\newcolumntype{C}[1]{>{\centering\let\newline\\\arraybackslash\hspace{0pt}}m{#1}}
\newcolumntype{R}[1]{>{\raggedleft\let\newline\\\arraybackslash\hspace{0pt}}m{#1}}
\newcommand\lengthMAE{3.4}

\usepackage[author={Hessam}]{pdfcomment}

\usepackage{xargs}  
\usepackage{textcmds} 

\definecolor{LightCyan}{rgb}{0.88,1,1}
\definecolor{LightGreen}{rgb}{0.88,1,0.88}
\definecolor{DarkGreen}{rgb}{0.0, 0.7, 0.1}
\definecolor{CommentColor}{rgb}{0.9, 0.3, 0.7}

\newcommand\Vtextvisiblespace[1][.3em]{%
  \mbox{\kern.06em\vrule height.3ex}%
  \vbox{\hrule width#1}%
  \hbox{\vrule height.3ex}}

\hypersetup{
    colorlinks,
	linkcolor=green,
    citecolor=green,
    urlcolor=green,
    filecolor=green
}

\begin{document}

\begin{frontmatter}
\title{Quantitative Error Prediction of Medical Image Registration using Regression Forests}

\cortext[cor1]{Corresponding author}
\author[LUMCadd]{Hessam Sokooti\corref{cor1}}

\ead{h.sokoot\_oskooyi@lumc.nl}
\author[Ankara_add]{Gorkem Saygili}
\author[Impadd]{Ben Glocker}
\author[LUMCadd,Delftadd]{Boudewijn P.F. Lelieveldt}
\author[LUMCadd,Delftadd]{Marius Staring}
\address[LUMCadd]{Leiden University Medical Center, Leiden, The Netherlands}
\address[Ankara_add]{Ankara University, Ankara, Turkey}
\address[Impadd]{Imperial College, London, United Kingdom}
\address[Delftadd]{Delft University of Technology, Delft, The Netherlands}

\begin{abstract}

Predicting registration error can be useful for evaluation of registration procedures, which is important for the adoption of registration techniques in the clinic. In addition, quantitative error prediction can be helpful in improving the registration quality. The task of predicting registration error is demanding due to the lack of a ground truth in medical images. This paper proposes a new automatic method to predict the registration error in a quantitative manner, and is applied to chest CT scans. A random regression forest is utilized to predict the registration error locally. The forest is built with features related to the transformation model and features related to the dissimilarity after registration. The forest is trained and tested using manually annotated corresponding points between pairs of chest CT scans in two experiments: SPREAD (trained and tested on SPREAD) and inter-database (including three databases SPREAD, DIR-Lab-4DCT and DIR-Lab-COPDgene). The results show that the mean absolute errors of regression are 1.07 $\pm$ 1.86 and \mbox{1.76 $\pm$ 2.59 mm} for the SPREAD and inter-database experiment, respectively. The overall accuracy of classification in three classes (correct, poor and wrong registration) is 90.7\% and 75.4\%, for SPREAD and inter-database respectively. The good performance of the proposed method enables important applications such as automatic quality control in large-scale image analysis.
\end{abstract}


\end{frontmatter}

\section{Introduction}

Image registration is the task of finding the optimal spatial transformation between two or more images. In most registration methods, no assessment of the registration quality is provided, and simply the result is returned.
Evaluation of the registration is devolved to human experts, which is very time-consuming and prone to inter-observer errors as well as human fatigue \citep{murphy2011evaluation}. Automatic quantitative error prediction of registration would decrease quality assessment time and can provide information about the registration uncertainty. 
Many medical pipelines are based on registered images and it is important to know the uncertainty of registration before continuing to a next phase in order to prevent accumulation of errors.
For example, in online adaptive radiotherapy daily contouring of the tumor and organs-at-risk can be performed with the help of image registration \citep{thornqvist2010propagation}. In this task, quality assessment (QA) is mandatory to ensure patient safety. In addition, the accumulation of delivered dose over several treatment fractions is also impacted by the quality of registration \citep{murphy2012method, tilly2013dose, veiga2015toward}. Registration quality therefore has to be checked before the treatment starts. Visualizing the error of registration can also be directly helpful in medical applications before making a clinical decision. \citet{smit2017pelvis} localized autonomic pelvic nerves by registering a pre-operative MRI scan to an atlas model that includes nerve information. These nerves are not visible in the MRI scans and are prone to be damaged during a surgical procedure. Utilizing registration uncertainty yield better visualization of the autonomic nerves. 

Refinement of registration is another important application of automatic error prediction.  \citet{muenzing2014dirboost} improved registration by focusing only on regions with high registration error and discarding pixels which are aligned correctly. Registration refinement can also be done with the feedback of human experts by manually adding several corresponding landmarks \citep{gunay2017semi}.

\citet{schlachter2016visualization} did a comprehensive study on visualization of registration quality with the help of three radiation oncologists on the DIR-Lab-COPDgene data, which has a slice thickness of \mbox{2.5 mm}. The [average, maximum] TRE of the landmarks that were rated to be of acceptable registration quality with the conventional visualization method (checkerboard visualization and color blended) was \mbox{[2.3, 6.9] mm}, while with the best visualization method (histogram intersection) \mbox{[1.8, 3.3] mm} was achieved.

A few methods have been proposed to detect the misalignment of a pair of images with the purpose to refine the registration result. \citet{rohde2003adaptive} proposed to use the gradient of the cost function to detect which region in the image pair is poorly registered and potentially can be improved. \hyphenation{Schnabel}
\citet{schnabel2001generic} suggested to refine the registration by increasing the number of registration parameters in regions with high local entropy, or with high local variation in the intensity or with relatively steep cost function. In another work, analyzing the shape of the cost function around each voxel was used to estimate the confidence of registration \citep{saygili2016confidence}. \citet{park2004adaptive} used normalized local mutual information to find poorly aligned regions in order to increase the number of registration parameters. \citet{forsberg2011improving} utilized the outer product of the intensity gradient as an uncertainty measure in multi-channel diffeomorphic Demons registration. Although the mentioned metrics can be used to improve the image registration, it has not been shown how these metrics are correlated with the image registration error.

Several methods exploit continuous probabilistic image registration by utilizing Bayesian inference to achieve an intrinsic transformation uncertainty measure \citep{risholm2013bayesian, simpson2015probabilistic}. However, it has been shown that there is no clear statistical correlation between transformation uncertainty and registration uncertainty \citep{luo2017misdirected}. The transformation and corresponding label (of a pair of images) are two random variables and it is not possible to quantify the uncertainty of the corresponding label by the summary statistics of the transformation. Another downside of these methods is that they can only be used for the specific paradigm of Bayesian registration. 

Some methods are based on the consistency of multiple registrations between a group of images \citep{datteri2012automatic, gass2015consistency}, but these methods cannot be used in pairwise registrations. 

In the stochastic approaches, \citet{kybic2010bootstrap} suggested to perform multiple registrations with random sampling of pixels with replacement. He found a correlation between the true registration error and the variation of the 2D translational parameters. The method was not extended to 3D and to nonrigid registration. \citet{hub2009stochastic} calculated the local mean square intensity difference multiple times by perturbing the B-spline grid. They showed that the maximum change of the dissimilarity metric in a local region is correlated with the registration error in that region. The drawback of this method is that it is not efficient in homogeneous areas \citep{hub2013estimation}. In a related work they showed that the variance of the final deformation vector field (DVF) is related to the registration error \citep{hub2013estimation}, using the Demons algorithm. However, to find large misalignment a large search region is needed.

In this paper, we turn our attention to methods capable of \emph{learning} the registration error allowing to take advantage of multiple features related to registration uncertainty within a single framework. \citet{muenzing2012supervised} casted the registration assessment task to a classification problem with three categories (wrong, poor and correct registrations). In their method, they mostly utilize intensity-based features, except for the determinant of the Jacobian of the transformation. Although their training samples consist of manually selected landmarks, later they showed that assessing registration in all regions is possible by interpolation \citep{muenzing2014dirboost}.

In our paper, instead of casting the uncertainty estimation task to a classification problem, we formulate it as a regression problem. To the best of our knowledge, in the field of continuous prediction of 3D registration error, \citet{lotfi2013improving} only tested their method on artificially deformed images. Recently \citet{eppenhof2017supervised} estimated the registration error by utilizing convolutional neural networks. Only preliminary results were available for synthetic 3D data.

We explore several features related to the uncertainty of the registration transformation as well as related to intensity. All features are calculated in physical units, i.e. mm, which makes the system independent of voxel size. Finally, features are combined by using regression forests. The proposed method is applied and evaluated on chest CT scans. This work is an extension of \cite{sokooti2016accuracy} with updated methodology and substantially extended evaluation.

\section{Methods}
\subsection{System overview}

\begin{figure*}[tb] 
	\centering
	\includegraphics[width=1\textwidth]{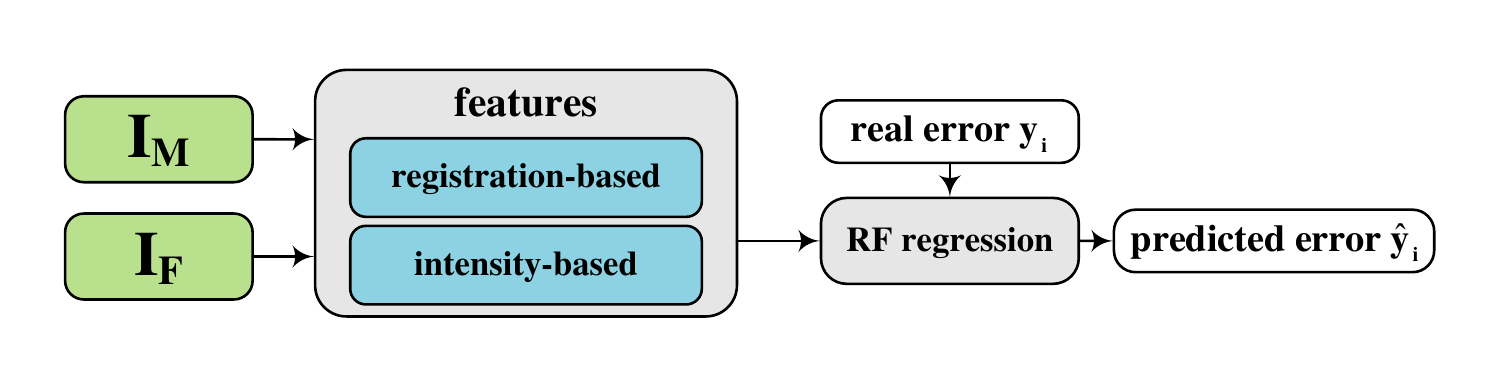}
	\caption{A block diagram of the proposed algorithm.}\label{fig:BD}
\end{figure*}

A block diagram of the proposed algorithm is shown in Fig. \ref{fig:BD}. The system has two inputs: a fixed image $I_{F}$ and a moving image $I_{M}$. Several registration-based and intensity-based features are generated. A regression forests (RF) is then trained from all features to estimate the registration error. 

The proposed system is trained to predict residual distances $y$ (registration errors) obtained from a set of semi-automatically established corresponding landmarks. During evaluation, the prediction result $\hat y$ is compared with errors obtained from an independent set of ground truth landmarks, using cross-validation. The proposed system therefore estimates registration errors in physical units, i.e. mm.  More information about the ground truth is available in Section \ref{title:Materials}. Details of the features are elaborated in Section \ref{hd:Features}.

\subsection{Registration}
Registration can be formulated as an optimization problem in which the cost function $\mathcal{C}$ is minimized with respect to $\bm{T}$:
\begin{align}
	\widehat{\bm{T}}=\text{arg}\, \underset{\bm{T}}{\text{min}} \, \mathcal{C} \big( \bm{T};I_F,I_M \big),
	\label{eq:reg}
\end{align}
where $\bm{T}$ denotes the transformation. The optimization is usually solved by an iterative method embedded in a multi-resolution setting. A registration can be initialized by an initial transform $\bm{T}^{\mathrm{ini}}$.

\subsection{Features and pooling} \label{hd:Features}
The features we used in our system, consist of several registration-based as well as intensity-based features. Some features are intrinsically capable to be calculated over differently sized local boxes, for others, a pool of features is created by computing local averages and maxima afterwards. The features used in this paper are listed in Table \ref{tb:Features}. We propose the following features:

\subsubsection{Registration-based features}
\begin{figure}[tb] 
	\centering
	\subfigure[]{
	\includegraphics[width=.98\columnwidth]{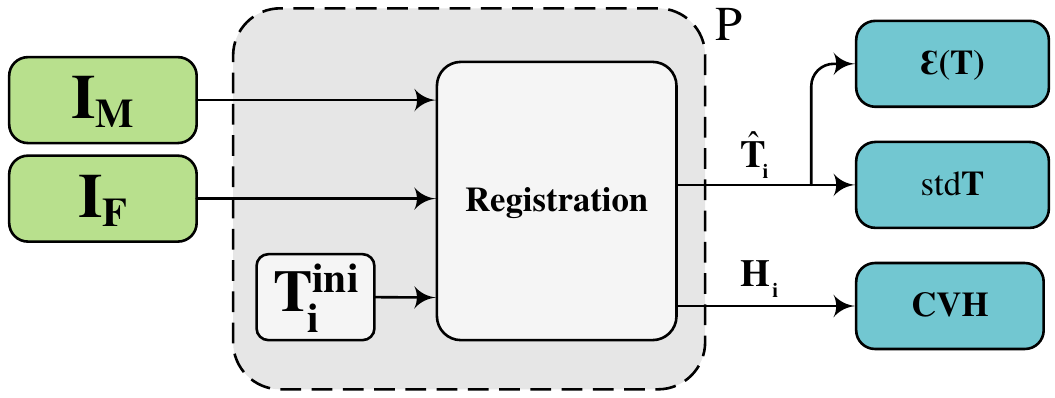}\label{fig:StdCVH}}
	\subfigure[]{\includegraphics[width=.98\columnwidth]{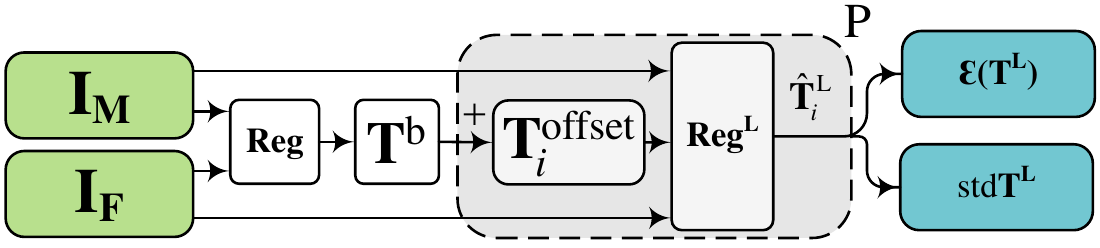} \label{fig:stdL}}
	\caption{Multiple registrations are performed to create
registration-based features. Either the initial transformation is varied, or the transformation after the base registration.}	
\end{figure}

\begin{figure}[tb] 
	\centering
	  \subfigure[$\std \bm{T}$]{\includegraphics[width=0.48\columnwidth, trim={0 0 0 0},clip]{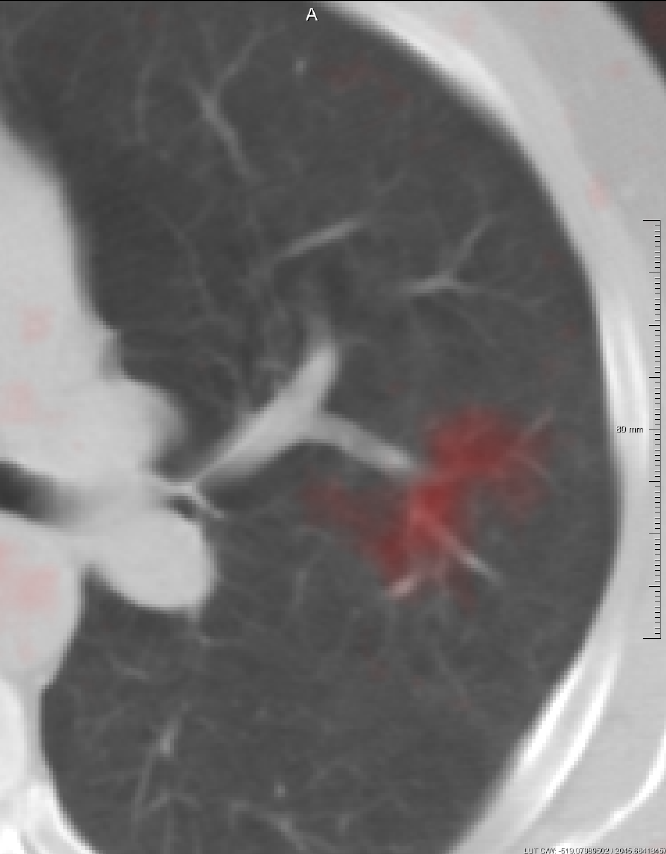}\label{fig:STD_Visual}}
  \subfigure[CVH]{\includegraphics[width=0.48\columnwidth, trim={0 0 0 0},clip]{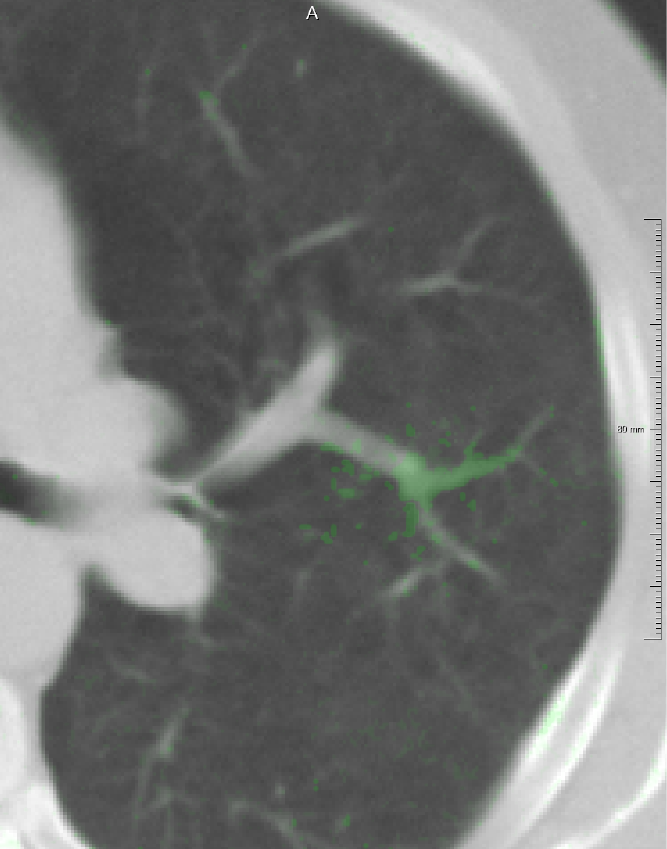}\label{fig:CVH_Visual}}   
	\caption{Visualization of $\std \bm{T}$ and CVH in a synthetically deformed image. The deformed image is created by a random deformation vector field which is smoothed by a Gaussian kernel similar to \protect{\cite{sokooti2017nonrigid}}.}
\end{figure}

\textbf{Variation of deformation vector field ({$\std{\bm{T}}$}):}
The final solution of an iterative optimization problem can be influenced by the initial parameters. If in a region the cost function has multiple local minima or is semi-flat, a slight change in the initial parameters can lead to a different solution. In contrast, in areas where the cost function is well-defined, variations in the initial state are expected to have much less effect on the final solution. A flow chart of the described feature is available in \mbox{Fig. \ref{fig:StdCVH}}. Given $P$ random initial transformations $\bm{T}^{\mathrm{ini}}_i$, $i \in \{1,\ldots,P\},$  that are used as initializations of the registration algorithm from Eq. (\ref{eq:reg}), the variation in the final transformation results $\widehat{\bm{T}}_i$ is a surrogate for the precision of the registration. We propose to use the standard deviation $\std{\bm{T}}$ of those final transformations as a feature:
\begin{align}
\overline{\bm{T}} &= \tfrac{1}{P} \sum \widehat{\bm{T}}_i, \\
\std{\bm{T}} &=  \sqrt{\tfrac{1}{P-1} \sum {\| \widehat{\bm{T}}_i - \overline{\bm{T}} \|}^2}.
\label{eq:std}
\end{align}
In this work, the initial transformations $\bm{T}^{\mathrm{ini}}_i$ are created by uniformly distributed offsets in the range $[-2,2]\,$mm to all B-spline coefficients. The offset range is chosen to be relatively small in comparison to the B-spline grid spacing in order to avoid unrealistic deformation. An example of $\std{\bm{T}}$ in a synthetically deformed image is given in \mbox{Fig. \ref{fig:STD_Visual}}.

Instead of perturbing the initial state of the registration, it is also possible to first perform the registration without any manipulated initial state, resulting in a transformation $\bm{T}^{\mathrm{b}}$ \citep{klein2009adaptive}. Then, random offsets $\bm{T}^{\mathrm{offset}}_{i}$ are added to $\bm{T}^{\mathrm{b}}$ after which another registration is performed, resulting in $\widehat{\bm{T}_i^\mathrm{L}}$. This is close to the work of \citet{hub2013estimation}, and approximately measures the concavity of the cost function. The feature $\std{\bm{T}^\mathrm{L}}$ is then derived akin to Eq. (\ref{eq:std}): 
\begin{align}
\overline{\bm{T}^\mathrm{L}} &= \tfrac{1}{P} \sum \widehat{\bm{T}_i^\mathrm{L}}, \\
\std{\bm{T}^\mathrm{L}} &=  \sqrt{\tfrac{1}{P-1} \sum {\| \widehat{\bm{T}_i^\mathrm{L}} - \overline{\bm{T}^\mathrm{L}} \|}^2}.
\label{eq:stdL}
\end{align}
It is expected that $\std{\bm{T}^\mathrm{L}}$ is small in regions
where the cost function is concave, as by adding small offsets $\bm{T}^{\mathrm{offset}}_{i}$ to the parameters, it can still move back to the previous optimal point. A flow chart of $\std{\bm{T}^\mathrm{L}}$ is shown in Fig \ref{fig:stdL}. $\std{\bm{T}^\mathrm{L}}$ is calculated using the same setting as $\std{\bm{T}}$, except that only one resolution is used.

If the difference between $\overline{\bm{T}}$ and $\bm{T}^{\mathrm{b}}$ is relatively large, regions indicating a small $\std{\bm{T}}$ are still potentially regions of low registration quality. We then consider the bias $\mathcal{E}(\bm{T})$ and $\mathcal{E}(\bm{T}^\mathrm{L})$ as complementary features to $\std{\bm{T}}$ and $\std{\bm{T}^\mathrm{L}}$ computed by:
\begin{equation}
	\begin{aligned}
\mathcal{E}(\bm{T}) &= \| \bm{T}^{\mathrm{b}} - \overline{\bm{T}} \|,\\ 
\mathcal{E}(\bm{T}^\mathrm{L}) &= \| \bm{T}^{\mathrm{b}} - \overline{\bm{T}^\mathrm{L}} \|.
\label{eq:bias}
	\end{aligned}
\end{equation}

\textbf{Coefficient of variation of joint histograms (CVH)}:
Multiple registration results can be used to extract additional information from the matched intensity patterns of the images. 
Given a fixed image $I_F$ and a registration sub-result $I_M(\bm{T}_i)$, we calculate their joint histogram $\mathrm{H}_i , \forall i$. For identical sub-registrations, all resulting joint histograms are equal. Variation in the joint histograms implies registration uncertainty as a surrogate for registration error. The coefficient of variation of the joint histograms is calculated by dividing the standard deviation of all joint histograms over the average, $\overline{\mathrm{H}}$, of them. This normalization is done to compensate for large differences between the elements of $\overline{\mathrm{H}}$. We obtain the CVH in histogram space as follows:
\begin{equation}
    \begin{aligned}\label{eq:CV}    
      \mathrm{CVH^{B \times B}} &= \frac{\std \mathrm{H}}{\overline{\mathrm{H}} + \epsilon},
    \end{aligned}
  \end{equation}
where B is the number of histogram bins, and $\epsilon$ a constant to avoid division by zero. In the experiments we set {$\epsilon$} to 5. The CVH$^{\mathrm{B} \times \mathrm{B}}$ in histogram space is subsequently transferred to the spatial domain, by assigning voxels $x$ with a particular intensity combination $\big(I_F(x), I_M(\bm{T}^\mathrm{b}(x))\big)$ the corresponding value from CVH$^{\mathrm{B} \times \mathrm{B}}$, resulting in the final CVH feature with size equal to the fixed image. Note that the CVH can be used in a multi-modality setting, like the previous features. An example of the CVH on a synthetically deformed image is given in Fig. \ref{fig:CVH_Visual}.

\textbf{Determinant of the Jacobian (Jac)}: Jac measures the relative local volume change. This can point to poor registration quality in case of very large ($\mathrm{Jac} \gg 1$) or very small ($\mathrm{Jac} \ll 1$) values, or discontinuous transformations in case of a negative value ($ \mathrm{Jac} < 0$). In the experiments, the determinant of the Jacobian of $\bm{T}^{\mathrm{b}}$ is used.

\subsubsection{Intensity-based features}
\begin{figure}[tb]
\centering
\subfigure[2D projection]{\includegraphics[width=.45\columnwidth ,trim ={0 0 0 0}]{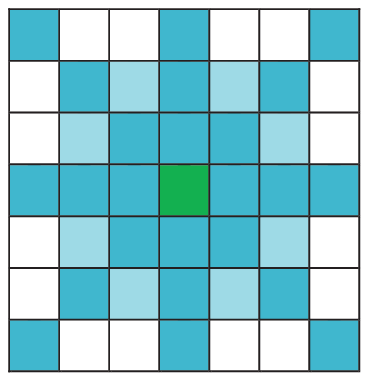}}
\subfigure[3D view]{\includegraphics[width=0.35\columnwidth , trim = 0 0 0 0]				{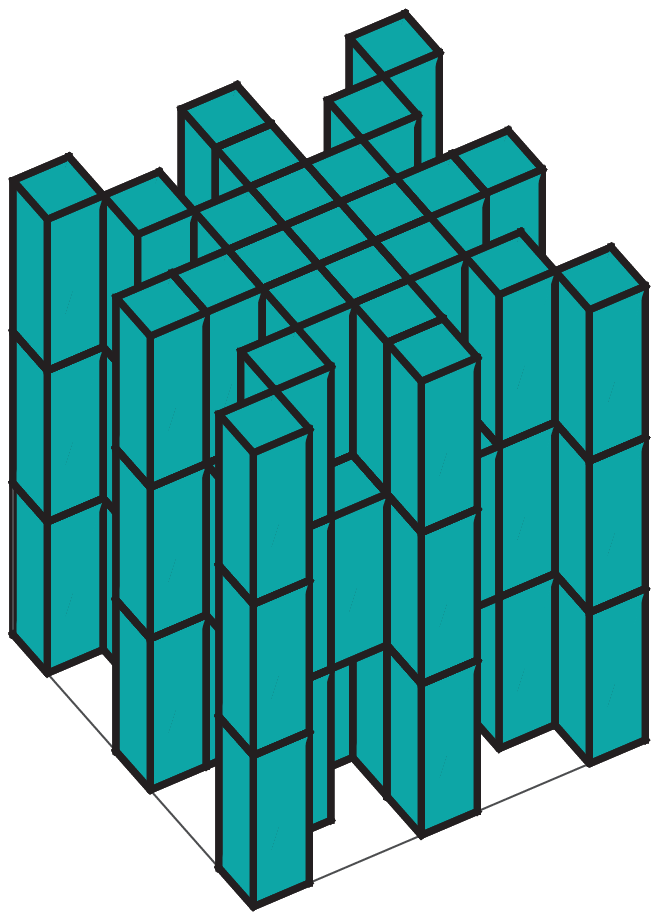}}
\caption{MIND search region. (a) The green cell indicates the center and darker blue cells indicate more accumulated cells in the projection view. }	
\label{fig:MIND3D}
\end{figure}

\textbf{MIND}: The Modality Independent Neighborhood Descriptor (MIND) was introduced by \citet{heinrich2012mind} in order to register multi-modal images. In this local self-similarity metric, a patch is considered to compare intensities between fixed and moving images. Finally, the sum of absolute differences between the MIND vector of $I_F$ and that of $I_M(\bm{T}^{\mathrm{b}})$ is computed. We calculate MIND with a sparse patch including 82 voxels inside a $[7 \times 7 \times 3]$ box, which is approximately physically isotropic for the data used in the experiments (see Fig. \ref{fig:MIND3D}).

\textbf{Local normalized mutual information}: Mutual information is used as an entropy-based similarity measure of two images. Similar to \cite{muenzing2012supervised} we use the following definitions for local normalized mutual information:
\begin{equation}
    \begin{aligned}\label{eq:NMI}    
      \mathrm{NMI} &= \frac{H(I_F)+H(I_M(\bm{T}^{\mathrm{b}}))}{H\Big(I_F,(I_M(\bm{T}^{\mathrm{b}})\Big)},\\      
      \mathrm{PMI} &= \frac{MI\Big(I_F,I_M(\bm{T}^{\mathrm{b}})\Big)}{min\Big\{H(I_F),H(I_M(\bm{T}^{\mathrm{b}}))\Big\}}.
    \end{aligned}
  \end{equation}
Both metrics are calculated over 8 differently sized boxes: [5, 10, 15, 20, 25, 30, 35, 40] mm. Two strategies for the selection of the number of bins are used, one uses a constant value $\mathrm{B_C}$, the other strategy depends on the number of samples $\mathrm{|B|}=log_2(n)+1$, in which $n$ is the number of samples in each box. The notations $\mathrm{NMIS}$ and $\mathrm{PMIS}$ indicate mutual information calculated with the latter strategy.  

\textbf{Modality-dependent features:} In addition to the modality-independent features from above, we consider the use of several modality-dependent features. In the experiments we assess their contributed value. Similar to \cite{muenzing2012supervised} the squared intensity difference (SID) and the gradient of intensity difference (GID) are computed using Gaussian (derivative) operators with standard deviations of [0.5, 1, 2, 4, 8, 16] mm. Normalized correlation (NC) is calculated within boxes of size [5, 10, 15, 20, 25, 30, 35, 40] mm akin to \cite{muenzing2012supervised}.

\subsubsection{Pooling} 

In order to reduce discontinuities and improve interaction with other features, the total set of features is increased by generating a pool from those mother features by calculating averages and maxima over them using differently sized boxes. The features MI, SID, GID and NC are inherently computed over differently sized local regions. The features $\mathrm{MIND}$, $\std\bm{T}$, $\std\bm{T}^\mathrm{L}$, CVH, $\mathcal{E}(\bm{T})$, $\mathcal{E}(\bm{T}^{\mathrm{L}})$ and Jac are calculated in a voxel-based fashion, and then pooled afterwards. Average and maximum pooling is performed with box sizes of {$[2, 5, 10, 15, 20, 25, 30, 35, 40]$} mm. As a result, for each feature we obtain a pool of 18 features: 9 from box averages and 9 from box maxima. The average-pooling is done efficiently by the help of integral images introduced by \citet{viola2004robust}. A list of the proposed mother features together with the number of derived features $N_f$ are given in Table \ref{tb:Features}.
\begin{table}[!tb]
	\centering
	\caption{An overview of the proposed features. Averages and maxima are taken over boxes of diameter [2, 5, 10, 15, 20, 25, 30, 35, 40] mm for the features: $\mathrm{MIND}$, $\std\bm{T}$, $\std\bm{T}^{\mathrm{L}}$, CVH, $\mathcal{E}(\bm{T})$, $\mathcal{E}(\bm{T}^{\mathrm{L}})$ and Jac. Mutual information measures are calculated in boxes of [5, 10, 15, 20, 25, 30, 35, 40] mm. SID and GID are computed using Gaussian derivatives with standard deviations in the range [0.5, 1, 2, 4, 8, 16] mm.
}
\resizebox{1\columnwidth}{!}{
\begin{tabular}{|lcl|}
\hline
      Feature   & $N_f$	&  		\\
  \hline		
	$\mathrm{MIND}$      	&18   &9 average boxes + 9 maxima boxes    \\
	MI      					&32   &NMI, NMIS, PMI, PMIS calculated over 8 boxes   \\
\hline
	$\std\bm{T}$      			&18   &9 average boxes + 9 maxima boxes    \\
	$\std\bm{T}^{\mathrm{L}}$ &18   &9 average boxes + 9 maxima boxes    \\
	CVH      					&18   &9 average boxes + 9 maxima boxes    \\
	$\mathcal{E}(\bm{T})$      	&18   &9 average boxes + 9 maxima boxes    \\
	$\mathcal{E}(\bm{T}^{\mathrm{L}})$      &18   &9 average boxes + 9 maxima boxes    \\
	Jac      					&18   &9 average boxes + 9 maxima boxes    \\
\hline
	NC      					&8    &calculated over 8 boxes    \\
	SID\&GID      				&12   &calculated over 6 sigma's   \\
	\hline	
	\end{tabular}}
	\label{tb:Features}	
\end{table}

\subsection{Regression forests}
Random forests were introduced by \citet{breiman2001random} by extending the idea of bagging. The forests consist of several weak learners (trees) which are combined in an efficient fashion. Each tree is started from a node and continues splitting until reaching certain criteria. In contrast to bagging, splitting is performed with a random subset of features which makes the training phase faster and reduces correlation between trees, consequently decreasing the forest error rate. The reason that we chose the random forest is that it can handle data without preprocessing. For instance rescaling of data, outlier removal and selection of features are not necessary in random forests. In addition, random forest are efficient to train and fast at runtime.

Random forests have the capability to calculate the importance of each feature with a little additional computation, which shows the contribution of each feature to the forest. Training of each tree is based on a bootstrap of all samples, and the so-called out-of-bootstrap samples $\Omega$ are used to compute the importance of a feature  $x_i$. Importance is then defined as the difference between the mean square error (MSE) before and after a permutation of this feature:

\begin{equation}
    \begin{aligned}
       \mathrm{Imp}(x_{i}) =  \frac{1}{N_{t}} \sum_{t=1}^{N_{t}} \bigg(\underset{j \in \Omega }{\mathrm{MSE}}\Big(\hat{y}_{\pi_{i}j},y_{j}\Big)-\underset{j \in \Omega }{\mathrm{MSE}}\Big(\hat{y_{j}},y_{j}\Big)\bigg),
    \end{aligned}
    \label{eq:RFimportance}
\end{equation}
where {$y_j$} is the real value, {$\hat{y_j}$} the predicted value from the regression, {$\hat{y}_{\pi_{i}j}$} the predicted value when permuting feature {$i$}, and $N_t$ the number of trees.

In this work, random forests are trained with different combinations of the proposed features (see Table \ref{tb:Features}). The dependent variable $y$ is 
the registration error in mm, which is described in Section \ref{title:Materials}.

\section{Experiments and results}
\label{Experiments}
\subsection{Materials and ground truth} \label{title:Materials} The SPREAD \citep{stolk2007progression} DIR-Lab-4DCT \citep{castillo2009framework} and DIR-Lab-COPDgene \citep{castillo2013reference} databases have been used in this study. In the SPREAD study, there are 21 pairs of 3D follow-up lung CT images. Each patient in this database has a baseline and a follow-up image (which is taken after 30 months) both in inhale phase. The age of the patients ranges from 49 to 78 years old. The average size of the images is $446\times 315 \times 129$ with an average voxel size of \mbox{$0.78\times 0.78 \times 2.50$ mm}. In each pair of images, about 100 well-distributed corresponding landmarks were previously selected \citep{staring2014towards} semi-automatically on distinctive locations \citep{murphy2011semi}.

From the DIR-Lab-4DCT data, five cases (4DCT1 to 4DCT5) are selected with each five phases between maximum inhalation and exhalation. The average image size is $256\times 256 \times 103$ with an average voxel size of \mbox{$1.10\times 1.10 \times 2.50$ mm}. Each scan has 75 corresponding landmarks annotated.
Ten cases with severe breathing disorders are available via the DIR-Lab-COPDgene database. The images are taken in inhale and exhale phases. In total, 300 landmarks are annotated. The average image size and the average voxel size are \mbox{$512\times 512 \times 120$} and \mbox{$0.64\times 0.64 \times 2.50$ mm}, \mbox{res}pectively.

\begin{figure}[!tb] 
	\centering
	
	\subfigure {\includegraphics[width=.997\columnwidth]	{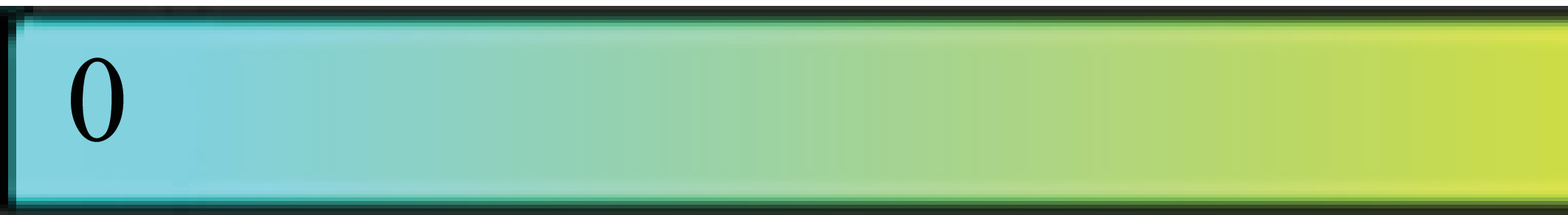}\label{fig:GT_a_GroundTruth} }
	{\vspace{-6mm}} 
	\addtocounter{subfigure}{-1}
	
	\subfigure[Ground truth]{\includegraphics[width=.49\columnwidth]	{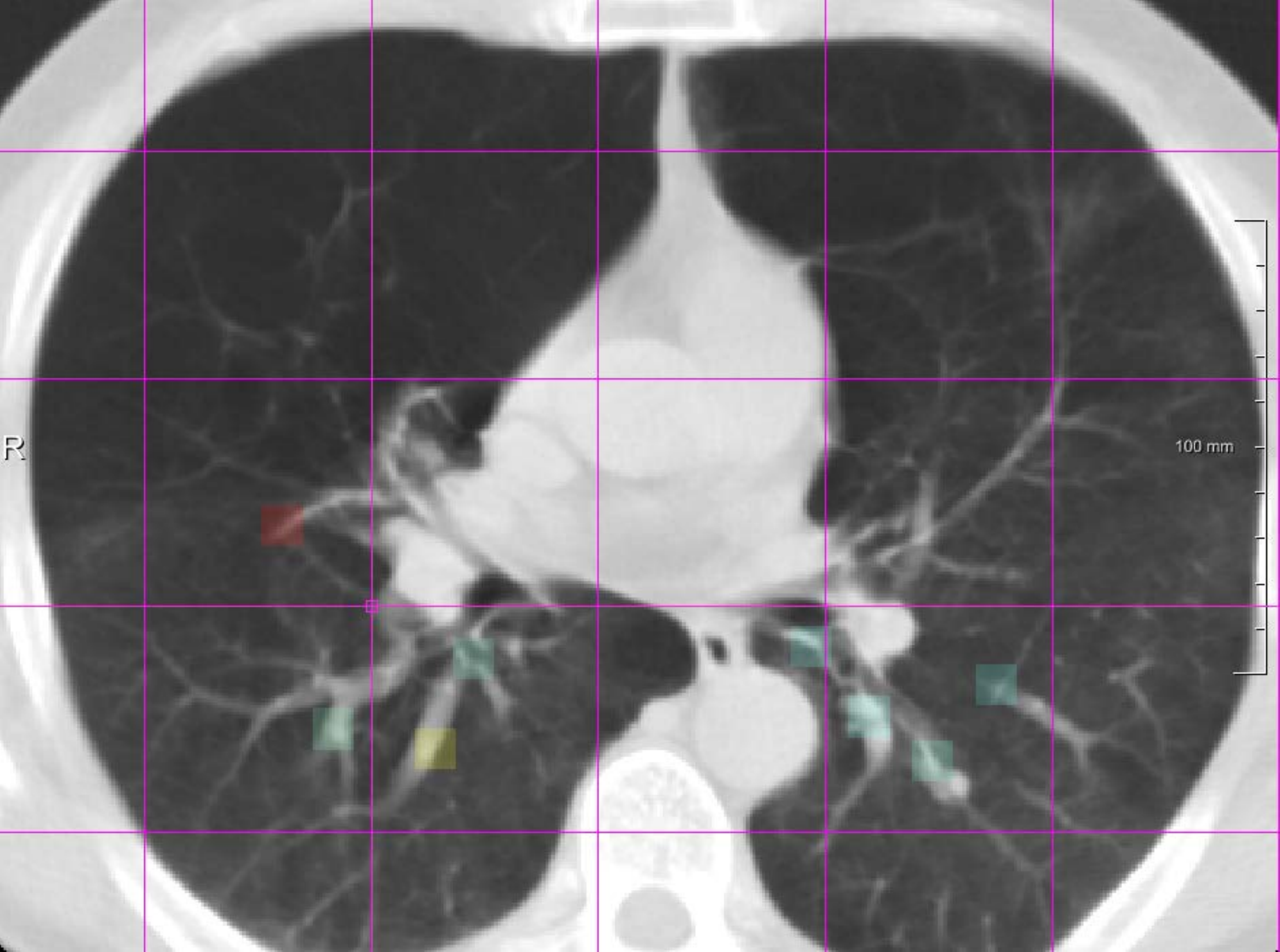}}
	\subfigure[Predicted error]{\includegraphics[width=.49\columnwidth]	{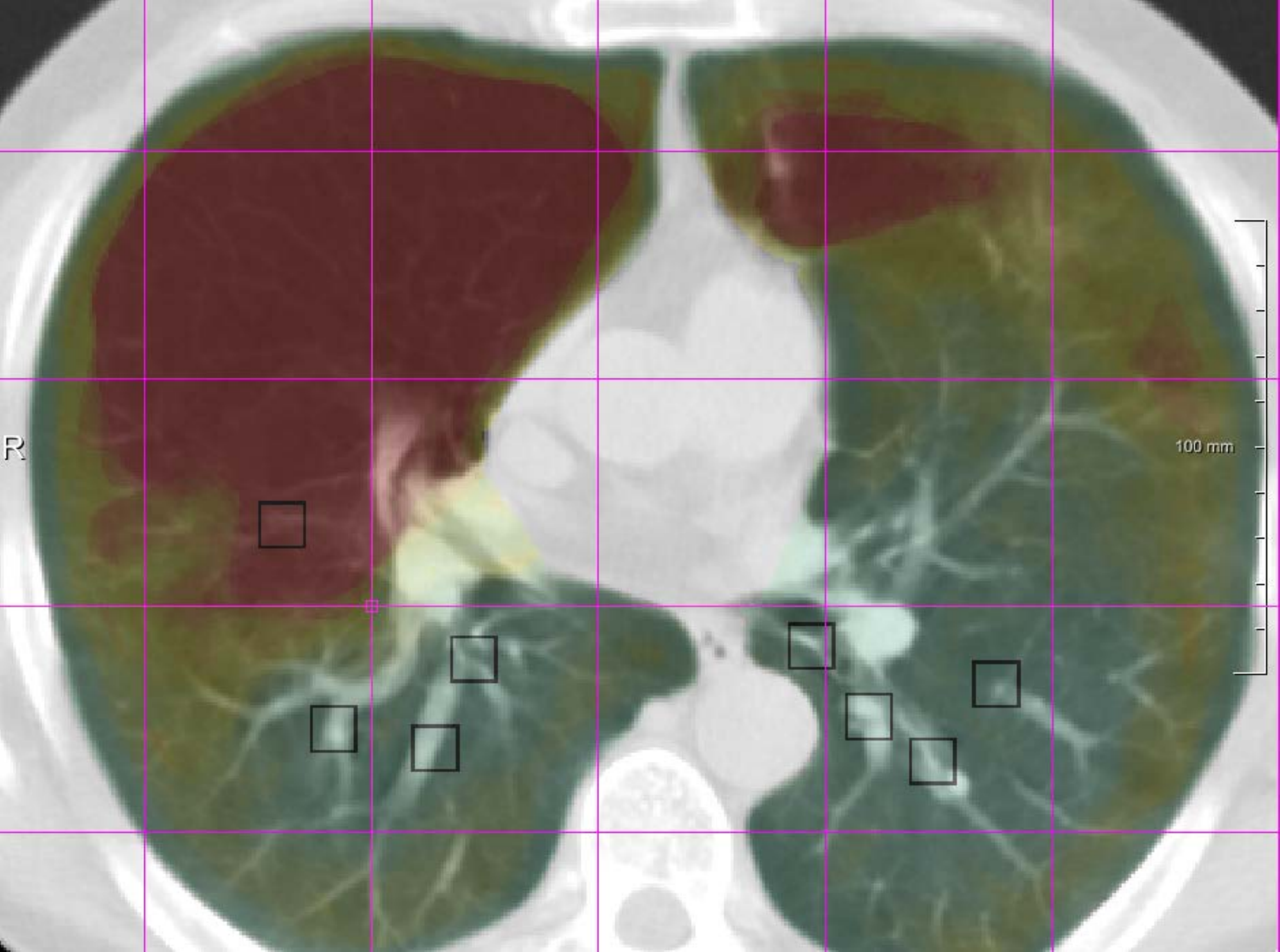}}
		
	\subfigure[Magnification of (a)]{\includegraphics[width=.49\columnwidth]	{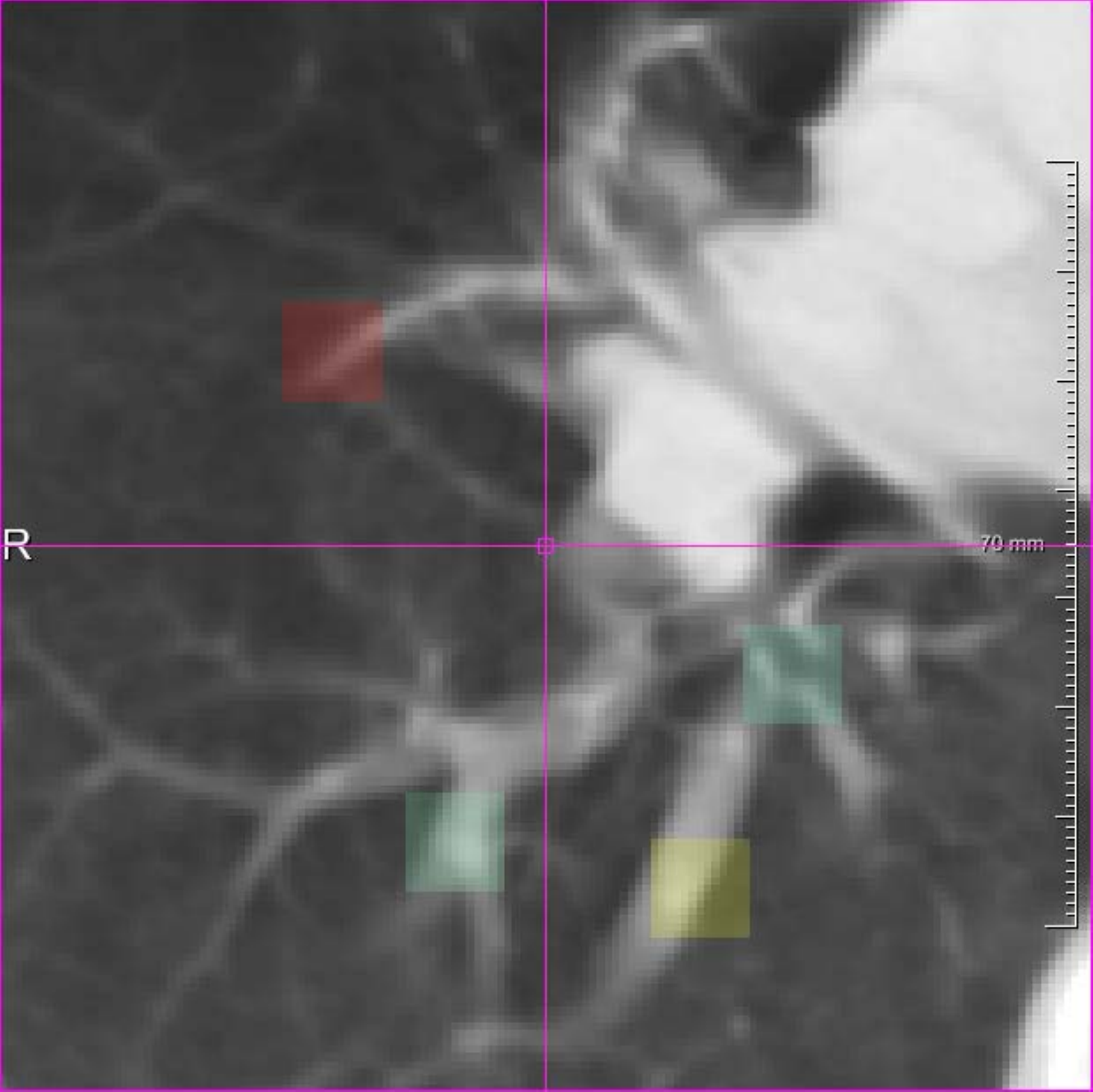}}
	\subfigure[Magnification of (b)]{\includegraphics[width=.49\columnwidth]	{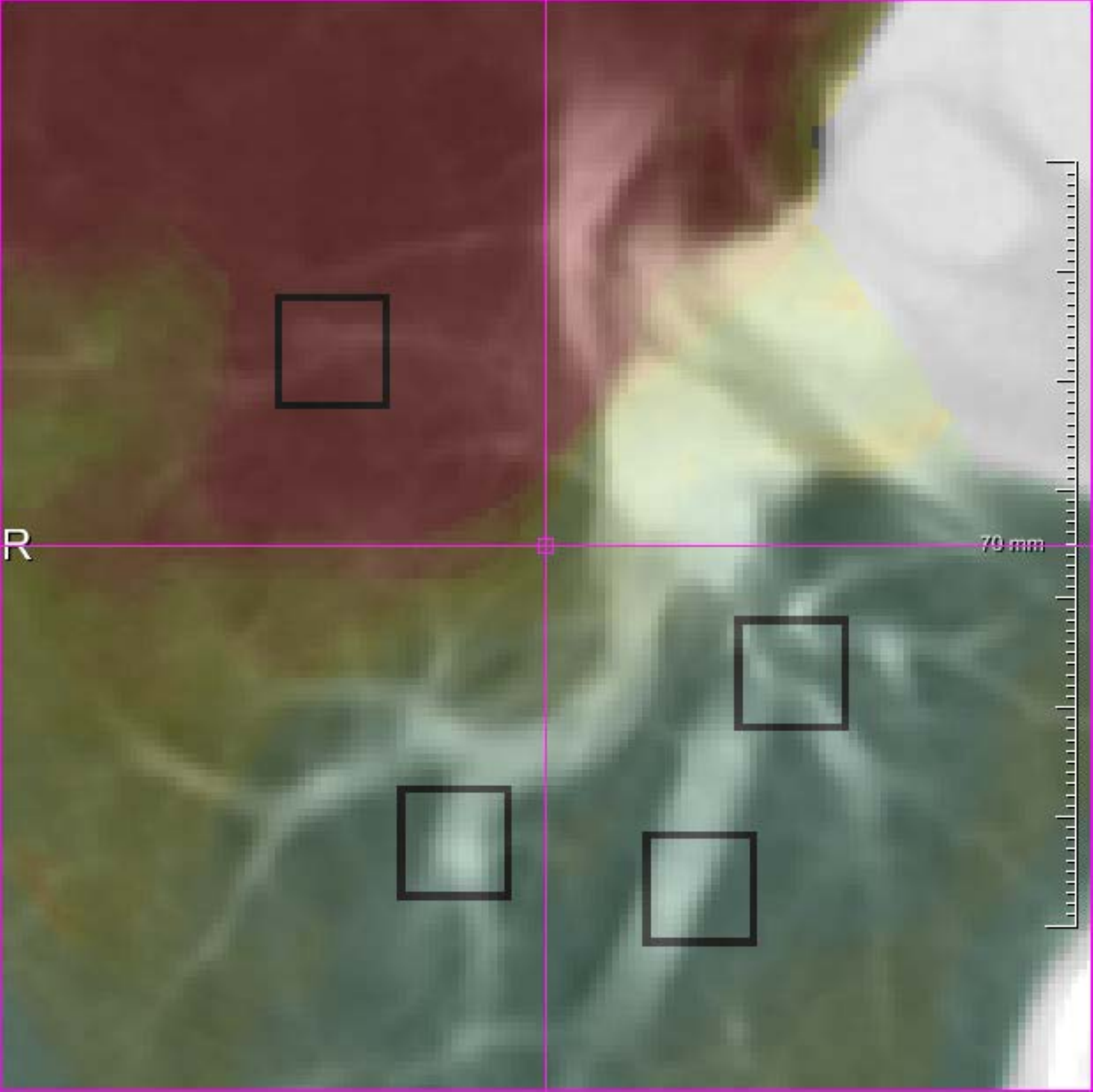}}
	\caption{Example data from the SPREAD dataset. The left column (a,c) shows the fixed image with the ground truth registration error overlaid in color. The square boxes around each landmark are given the same error as the error at the landmark. The right column (b,d) shows the moving image after registration with the registration error predicted by the proposed method overlaid in color. (c) and (d) are zoomed in versions of (a) and (b).}	
\label{fig:GT}
\end{figure}

Accuracy of the registration can be defined as the residual Euclidean distance after registration between the corresponding landmarks:
\begin{align}
y =  \|\bm{T}^{\mathrm{b}}(\bm{x}_F) - \bm{x}_M \|_2,
\label{eq:TRE}
\end{align}
with $\bm{x}_F$ and $\bm{x}_M$ the corresponding landmark locations. Based on the idea that the registration error is smooth, we include voxels from a small local neighborhood around the landmarks to increase the total set of available landmarks. In this small neighborhood we assume that the registration error is equal to the error at the center of the neighborhood. This assumption seems reasonable for smooth transformations and within a small region. The neighborhood size is chosen as {$10\times10\times7.5 \;\mathrm{mm}$}, which is approximately equivalent to the final grid spacing of the B-spline registration (see Fig. \ref{fig:GT_a_GroundTruth}).

The core software is written in Python. The feature pooling is performed with a C++ program \citep{glocker2014robust} and the regression forest is calculated with the help of the Scikit-learn package \citep{scikit-learn}. All registrations are performed by \texttt{elastix} \citep{klein2010elastix}. Detailed registration setting can be found in the \texttt{elastix} parameter file database (\href{http://elastix.bigr.nl/wiki/index.php/Parameter_file_database}{elastix.isi.uu.nl}, {entry par0049). The code is publicly available via \href{https://github.com/hsokooti/regun}{github.com/hsokooti/regun}.

\subsection{Evaluation measures} In the SPREAD database, we employ 10 cross-validations by randomly splitting the data in 15 image pairs for training and the remaining 6 pairs for testing. To evaluate the regression performance, the mean absolute error (MAE) of the real registration error $y_i$ and the estimated one $\hat{y}_i$ is calculated over the neighborhood of the landmarks by:
\begin{equation}
	\begin{aligned} 
       \mathrm{MAE} &=  \frac{1}{N} \sum_{i=1}^{N} |\hat{y}_i - y_i |. \\       
    \end{aligned} 
    \label{eq:RegError}    
\end{equation}
To further detail the regression performance, the MAE is subdivided into three categories: $\mathrm{MAE_c}$, $\mathrm{MAE_p}$ and $\mathrm{MAE_w}$ with $y$ in $[0,3)$, $[3,6)$ and \mbox{$[6,\infty)$ mm}, corresponding to correct, poor and wrong registration, similar to \citet{muenzing2012supervised}. We then do the same for $\hat{y}_i$, and report the accuracy and F1 score for classifying the registration error in these three categories.

\subsection{Parameter selection}
The RF is trained using 100 trees with a maximum tree depth of 9, while at least 5 samples remain in the leaf nodes. At each splitting node, $m$ features are randomly selected. We set $m$ to the square root of the total number of features in that experiment, which performed slightly better than $m= \text{(number of features)} / 3$ \citep{liaw2002classification}. The total number of registrations $P$ is chosen as 20 to ensure that the estimation of $\std\bm{T}$ does not change considerably when increasing the number of registrations \citep{sokooti2016accuracy}.

\subsection{Reference registration error set}

For the SPREAD and the DIR-Lab-4DCT study, registrations are based on free-form deformations by B-splines \citep{rueckert1999nonrigid}. The cost function is mutual information, which is optimized by adaptive stochastic gradient descent. We used three resolutions with a final B-spline grid spacing of \mbox{{$[10,10,10]$} mm}. We collect samples by performing four different registrations using 20, 100, 500 and 2000 iterations, respectively. All other registration settings remain the same in these registrations. By varying the number of iterations we increase the variation in the samples, as well as the training size. Table \ref{tb:Composition} gives the distribution of reference registration errors in each database. As expected, increasing the number of iterations shifts the distribution towards the ``correct" registration category. The maximum registration error is \mbox{81.8 mm} in the SPREAD database, \mbox{17.6 mm} in the DIR-Lab-4DCT database.

Since the a priori distribution of registration errors is imbalanced, with much more samples in the ``correct" category, we perform the following balancing step during training. For landmarks that fall in the category ``correct", we only add samples from a smaller neighborhood of \mbox{{$5\times5\times2.5 \;\mathrm{mm}$}} instead of the \mbox{{$10\times10\times7.5 \;\mathrm{mm}$}} neighborhoods used for landmarks in the categories ``poor" and ``wrong". The distribution of reference registration errors of the training samples is shown in Table \ref{tb:CompositionTraining}. 

For the DIR-Lab-COPDgene study, more advanced settings of the registration are used. In this experiment, samples are taken only on the landmark locations. More details are given in Section \ref{sec:inter-database}. The maximum registration error in this data is \mbox{31.5 mm}.

\begin{table*}[tb]
	\centering 
	\caption{Distribution of the reference registration errors in each database, used during testing.}
	\begin{tabular}{L{4cm}ccccccc}
\hline
	Database-iters & correct &  &poor &  &wrong  & &  total\\
	\hline
SPREAD 20 &$848789$ &$(84.1\%)$ &$102837$ &$(10.2\%)$ &$58059$ &$(5.8\%)$ &$1009685$ \\
SPREAD 100  &$904796$ &$(89.6\%)$ &$66467$ &$(6.6\%)$ &$38422$ &$(3.8\%)$ &$1009685$ \\
SPREAD 500  &$925840$ &$(91.7\%)$ &$51910$ &$(5.1\%)$ &$31935$ &$(3.2\%)$ &$1009685$ \\
SPREAD 2000  &$935676$ &$(92.7\%)$ &$46170$ &$(4.6\%)$ &$27839$ &$(2.8\%)$ &$1009685$ \\
SPREAD together &$3615101$ &$(89.5\%)$ &$267384$ &$(6.6\%)$ &$156255$ &$(3.9\%)$ &$4038740${\vspace{2mm}} \\
DIR-Lab-4DCT 20  &$521481$ &$(84.5\%)$ &$71282$ &$(11.5\%)$ &$24543$ &$(4.0\%)$ &$617306$ \\
DIR-Lab-4DCT 100  &$540989$ &$(87.6\%)$ &$61131$ &$(9.9\%)$ &$15186$ &$(2.5\%)$ &$617306$ \\
DIR-Lab-4DCT 500  &$553757$ &$(89.7\%)$ &$53067$ &$(8.6\%)$ &$10482$ &$(1.7\%)$ &$617306$ \\
DIR-Lab-4DCT 2000  &$561909$ &$(91.0\%)$ &$46679$ &$(7.6\%)$ &$8718$ &$(1.4\%)$ &$617306$ \\
DIR-Lab-4DCT together &$2178136$ &$(88.2\%)$ &$232159$ &$(9.4\%)$ &$58929$ &$(2.4\%)$ &$2469224$ {\vspace{2mm}}\\
DIR-Lab-COPD ANTsBSplineSyN &2643 &$(88.1\%)$ &184 &$(6.1\%)$ &173 &$(5.8\%)$ &3000 \\

DIR-Lab-COPD \texttt{elastix-advanced} &2420 &$(80.7\%)$ &259 &$(8.6\%)$ &321 &$(10.7\%)$ &3000 \\
	\hline		
	\end{tabular}
	\label{tb:Composition}	
\end{table*}

\begin{table*}[tb]
	\centering 
	\caption{Distribution of the reference registration errors, used during training.}
	\begin{tabular}{L{4cm}ccccccc}
\hline
	Database & correct &  &poor &  &wrong  & &  total\\
	\hline
SPREAD together &$589854$ &$(58.0\%)$ &$270523$ &$(26.6\%)$ &$156881$ &$(15.4\%)$ &$1017258$ \\

DIR-Lab-4DCT together &$328055$ &$(53.0\%)$ &$232499$ &$(37.5\%)$ &$58929$ &$(9.5\%)$ &$619483$ \\
	\hline	
	\vspace{1mm}
	\end{tabular}
	\label{tb:CompositionTraining}	
\end{table*}

\subsection{Experiments} \label{SpCross}
\subsubsection{Single feature performance in SPREAD}
 The proposed features are described in Section \ref{hd:Features} and summarized in Table \ref{tb:Features}. To investigate the strength of the individual features, we trained the random forest with only a single feature with pooling. By comparing the MAE results in Table \ref{tb:SpreadFeatureResult}, it can be seen that MIND, $\std\bm{T}^{\mathrm{L}}$ and SID$\&$GID are the best single features in the categories Intensity, Registration and Modality-dependent, respectively.

\begin{table*}[!tb]
	\centering
	\caption{Regression results for single features on the SPREAD database. The columns indicate the number of features ($N_f$), the mean absolute error (MAE), the accuracy (Acc) and the F1 score. The sub-indices c, p and w correspond to correct [0,3), poor [3,6) and wrong \mbox{[6, $\infty$) mm} classes, respectively. }
	\resizebox{\textwidth}{!}{
	\begin{tabular}{|L{\lengthMAE cm}|c|c|c|c|c|c|c|c|c|}
\hline
         &$N_f$ & MAE	& MAE\textsubscript{c} & MAE\textsubscript{p}	& MAE\textsubscript{w}  & \hspace{.6mm} Acc\hspace{.6mm}  & F1\textsubscript{c} & F1\textsubscript{p} & F1\textsubscript{w}\\
  \hline
$\mathrm{MIND}$ & 18  &$1.10\pm1.97$  &$0.76\pm0.72$  &$1.59\pm1.39$  &$6.50\pm5.88$  &$89.8$  &$94.9$  &$34.1$  &$83.0$  \\
MI & 32  &$1.20\pm1.88$  &$0.89\pm0.71$  &$1.53\pm1.14$  &$6.30\pm5.58$  &$87.9$  &$93.9$  &$30.1$  &$79.9$  \\
$\std\bm{T}$ & 18  &$1.59\pm2.79$  &$1.15\pm1.78$  &$2.98\pm4.06$  &$7.60\pm6.12$  &$85.5$  &$92.7$  &$22.4$  &$64.4$  \\
$\std\bm{T}^{\mathrm{L}}$ & 18  &$1.51\pm2.40$  &$1.11\pm1.34$  &$2.49\pm3.05$  &$7.32\pm5.79$  &$86.7$  &$93.4$  &$18.3$  &$70.7$  \\
CVH & 18  &$1.93\pm3.29$  &$1.49\pm2.22$  &$1.82\pm2.00$  &$9.80\pm7.19$  &$75.2$  &$87.2$  &$16.9$  &$37.0$  \\
$\mathcal{E}(\bm{T})$ & 18  &$2.00\pm2.80$  &$1.61\pm1.76$  &$2.18\pm3.12$  &$8.52\pm6.48$  &$69.8$  &$82.8$  &$17.0$  &$43.5$  \\
$\mathcal{E}(\bm{T}^{\mathrm{L}})$ & 18  &$1.68\pm2.85$  &$1.19\pm1.71$  &$3.19\pm3.28$  &$8.34\pm6.74$  &$84.4$  &$92.6$  &$11.7$  &$54.8$  \\
Jac & 18  &$2.15\pm3.15$  &$1.72\pm1.90$  &$1.91\pm2.27$  &$10.03\pm6.97$  &$68.2$  &$83.7$  &$13.0$  &$31.4$  \\
NC & 8  &$1.38\pm2.89$  &$0.90\pm0.71$  &$1.70\pm1.68$  &$9.41\pm9.15$  &$88.2$  &$94.3$  &$28.5$  &$77.0$  \\
SID\&GID & 12  &$1.30\pm2.12$  &$0.94\pm0.90$  &$1.82\pm1.63$  &$6.95\pm6.02$  &$89.9$  &$95.1$  &$24.9$  &$74.3$  \\
	\hline	
	\end{tabular}}
	\label{tb:SpreadFeatureResult}	
\end{table*}
 
\subsubsection{Combined features performance} \label{title:exp:combined}
Instead of using only a single feature, several combinations of features are used to build the RFs: 

\begin{itemize}

  	\item \textbf{Intensity:} Combination of all modality-independent intensity features: $\mathrm{MIND}$ and MI (50 features).
  	\item \textbf{Registration:} Combination of all registration features: $\std\bm{T}$, $\std\bm{T}^{\mathrm{L}}$, CVH, $\mathcal{E}(\bm{T})$, $\mathcal{E}(\bm{T}^{\mathrm{L}})$ and Jac (108 features).

	\item \textbf{Combined:} Combination of both intensity and registration features (158 features).
\end{itemize}

All results are available in Table \ref{tb:SpreadResult}. By combining features from both the registration and modality-independent intensity category, improvements were obtained in all evaluation measures.

The result of the regression with combined features is detailed in Fig. \ref{fig:RegSort}(a), which shows the real error (solid blue line) against the predicted error, sorted from small to large. In Fig. \ref{fig:RegSort}(b) we grouped the real errors in the three categories, each category showing a box-plot of the predicted errors. Intuitively, a smaller overlap between the boxes represents a better regression.

\begin{table*}[!tb]
	\centering
	\caption{Regression results for groups of features on the SPREAD database. The columns indicate the number of features ($N_f$), the mean absolute error (MAE), the accuracy (Acc) and the F1 score. The sub-indices c, p and w correspond to correct [0,3), poor [3,6) and wrong \mbox{[6, $\infty$) mm} classes, respectively. MD, NN and LR stands for modality dependent, neural networks and linear regression, respectively.}
	\resizebox{\textwidth}{!}{
	\begin{tabular}{|L{\lengthMAE cm}|c|c|c|c|c|c|c|c|c|}
\hline
         &$N_f$ & MAE	& MAE\textsubscript{c}	& MAE\textsubscript{p}	& MAE\textsubscript{w}  & \hspace{.6mm}Acc\hspace{.6mm}  & F1\textsubscript{c} & F1\textsubscript{p} & F1\textsubscript{w}\\
  \hline
Intensity & 50  &$1.09\pm1.88$  &$0.77\pm0.68$  &$1.49\pm1.26$  &$6.20\pm5.68$  &$90.3$  &$95.1$  &$35.7$  &$83.6$  \\
Registration & 108  &$1.32\pm2.35$  &$0.90\pm1.04$  &$2.10\pm2.71$  &$7.76\pm6.01$  &$90.0$  &$95.1$  &$31.5$  &$78.4$  \\
Combined & 158  &$1.07\pm1.86$  &$0.76\pm0.65$  &$1.47\pm1.22$  &$6.12\pm5.64$  &$90.7$  &$95.4$  &$38.1$  &$84.4$  \\
Combined-no pooling & 8  &$1.24\pm2.22$  &$0.85\pm0.73$  &$1.72\pm1.64$  &$7.39\pm6.62$  &$89.4$  &$94.8$  &$32.6$  &$79.1$  \\
Combined+MD & 178  &$1.07\pm1.83$  &$0.76\pm0.65$  &$1.46\pm1.20$  &$5.95\pm5.59$  &$90.7$  &$95.4$  &$38.3$  &$84.5$  \\
	\hline	
	Combined (LR) &158    &$1.86\pm2.03$  &$1.58\pm1.34$  &$2.47\pm2.21$  &$6.12\pm4.97$  &$77.3$  &$87.3$  &$17.0$  &$67.6$  \\
	Combined (NN) &158         &$1.13\pm2.07$  &$0.74\pm0.70$  &$1.81\pm1.67$  &$7.08\pm5.88$  &$89.8$  &$95.0$  &$31.2$  &$79.6$  \\
	\hline	
	\end{tabular}}
	\label{tb:SpreadResult}	
\end{table*}
\begin{figure*}[!tb] 
	\centering
	\subfigure[]{\includegraphics[width=.63\textwidth,trim={52 0 40 15},clip]{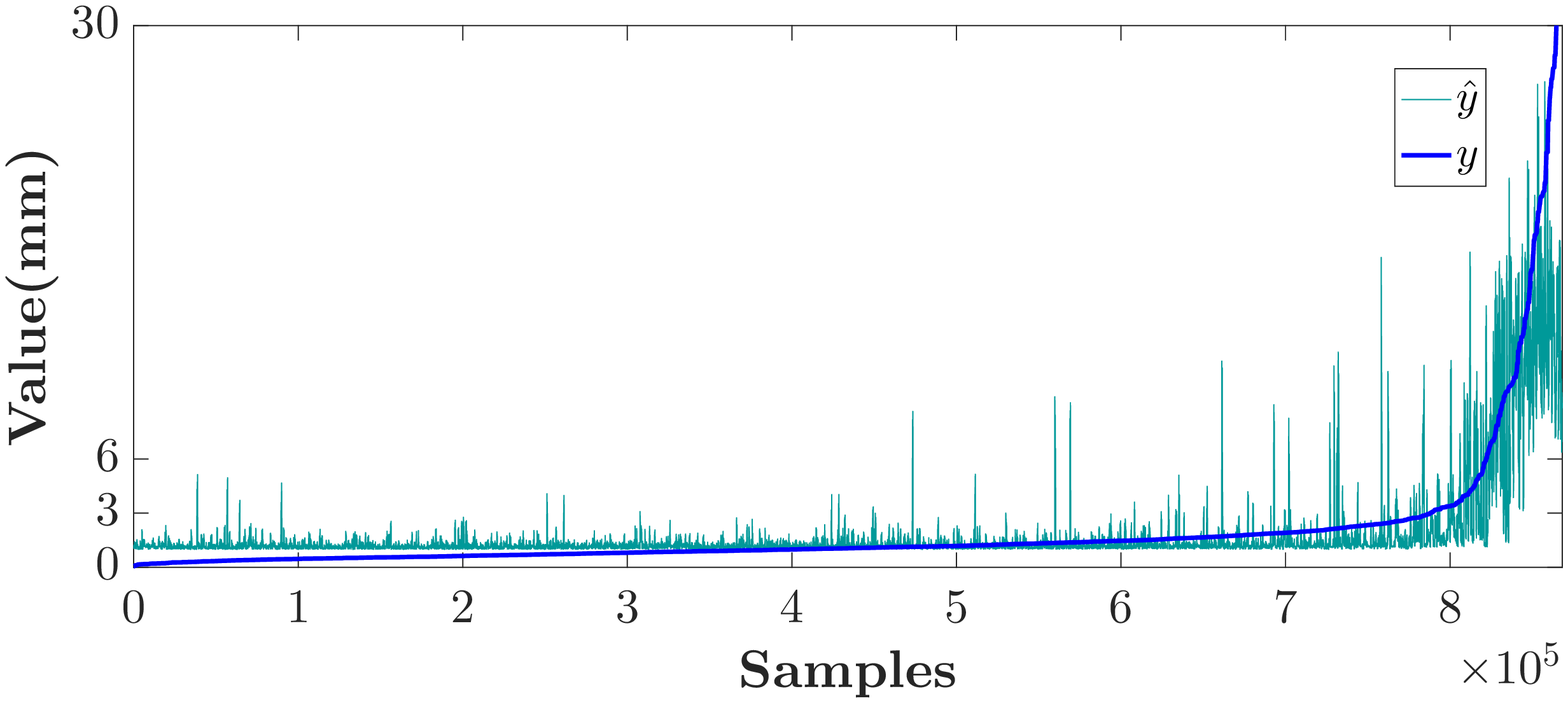}}
	\subfigure[]{\includegraphics[width=.35\textwidth,trim={2 0 27 13},clip]{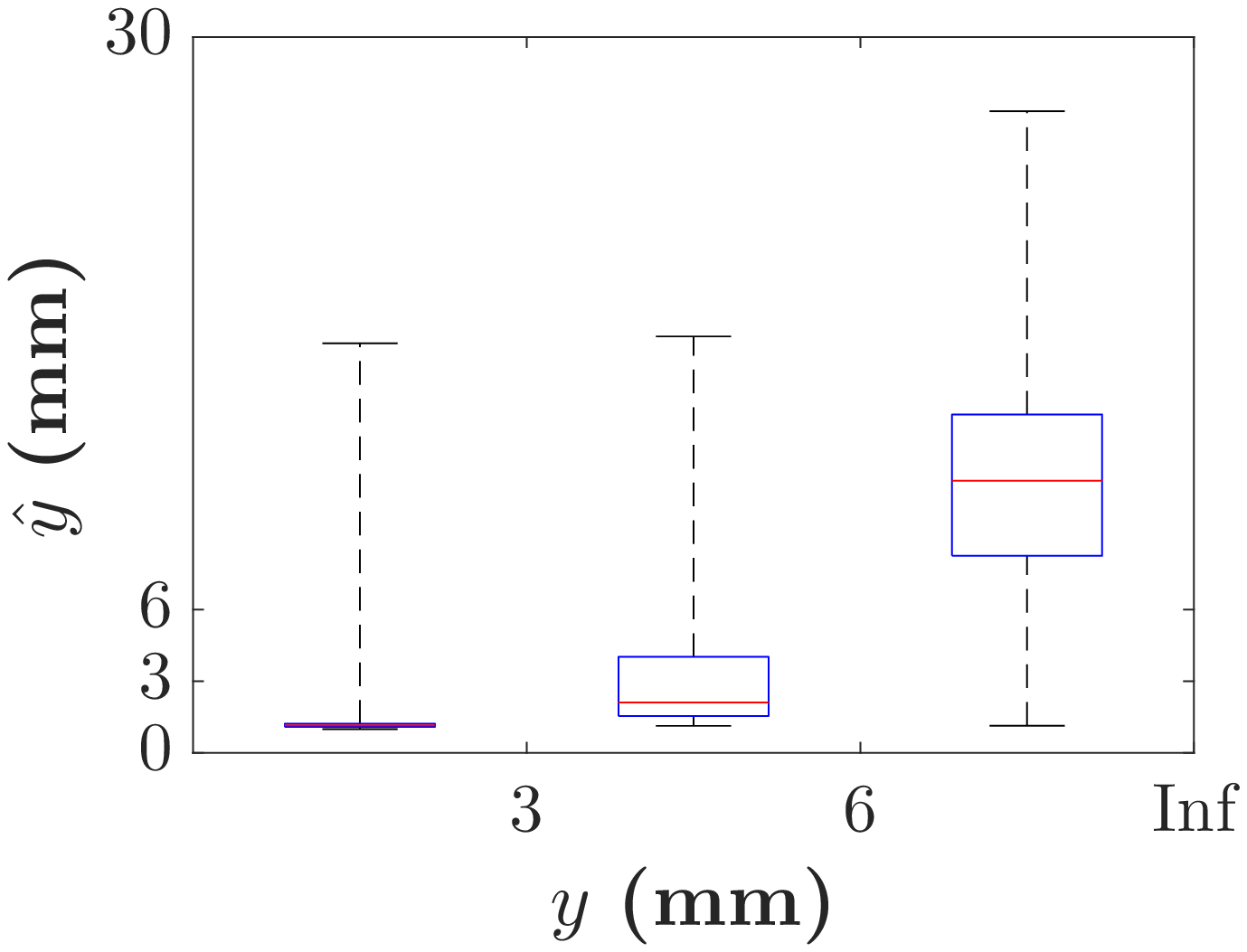}}   	
	\caption{Real ($y$) vs predicted registration error ($\hat y$) for Combined features in the SPREAD database. (a) The real error (solid blue line $y$) against the predicted error ($\hat y$), sorted from small to large. In (b) we grouped the real errors in the three categories, each category showing a box-plot of the predicted errors.}	
	\label{fig:RegSort}	
\end{figure*}

\subsubsection{Including modality-dependent features}
We consider adding the combination of three modality-dependent features to the combined feature set (Combined+MD): NC, SID and GID. In both databases, if we add the modality-dependent features (see Table \ref{tb:SpreadResult}), negligible differences are observed. Therefore, to keep the feature set small and modality-independent, we select the ``combined features" class without the  modality-dependent features as the final system in the remainder of this paper.

\subsubsection{The effect of pooling}
To examine the effect of pooling, we perform an experiment without pooling on the combined feature set. We only calculate PMIS within a box size of 15 mm in this experiment. From Table \ref{tb:SpreadResult} the benefit of pooling can be observed.

\subsubsection{Alternative regression methods}
In this section, we compare RF regression with linear regression (LR) and neural networks (NN). Feature normalization is done for both regressors. We utilized neural networks with three hidden layers of 1024, 512 and 256 units each. ReLU is used as an activation function and Huber is utilized as a loss function. Table \ref{tb:SpreadResult} gives the results of these experiments. The performance of neural networks is on par with random forests. However, the results of linear regression are not comparable to that of random forests, both in MAE and accuracy.

\subsubsection{Feature importance}

The feature importance, see Eq. (\ref{eq:RFimportance}), is displayed in Fig. \ref{fig:Importance}. It shows that MIND and MI  are the features contributing most to the RF performance, followed by $\std{\bm{T}}$, $\std\bm{T}^{\mathrm{L}}$ and CVH. 

The feature importance using a different number of iterations is shown in Fig. \ref{fig:ImportanceIterations}. The contribution of all intensity features stay the same in all experiments, while some of the registration features contribute differently with respect to the number of iterations. For instance, the importance of $\std\bm{T}$ and CVH increase with increasing the number of iterations. The features $\std\bm{T}^{\mathrm{L}}$ and $\mathcal{E}(\bm{T}^{\mathrm{L}})$ play important roles when the number of iterations is not enough for registration convergence. 

\begin{figure*}[!tb] 
	\centering	
	\subfigure{\includegraphics[width=1\textwidth,trim={80 0 100 0cm},clip]{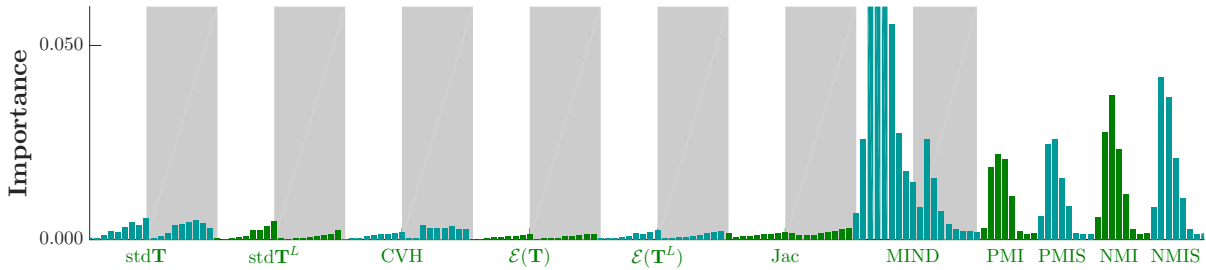}}
	\caption{Feature importance of the SPREAD combined experiment. White areas correspond to box averages, while shaded areas correspond to box maxima.}
    \label{fig:Importance}
\end{figure*}
\begin{figure*}[!tb] 
	\centering	
	\subfigure[SPREAD 20 iterations]{\includegraphics[width=1\textwidth,trim={80 0 100 0cm},clip]{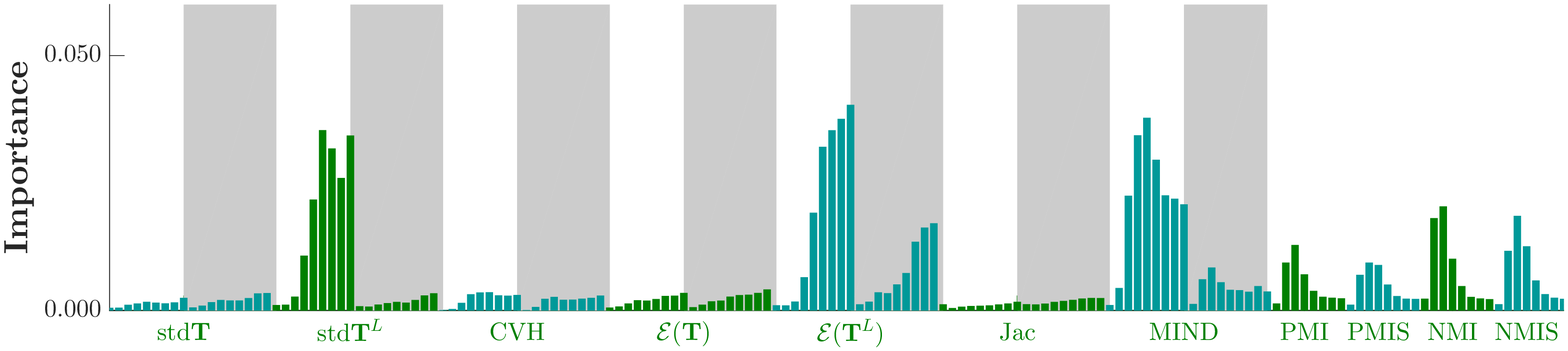} \label{fig:Imp:Sp20itr}}
	\subfigure[SPREAD 100 iterations]{\includegraphics[width=1\textwidth,trim={80 0 100 0cm},clip]{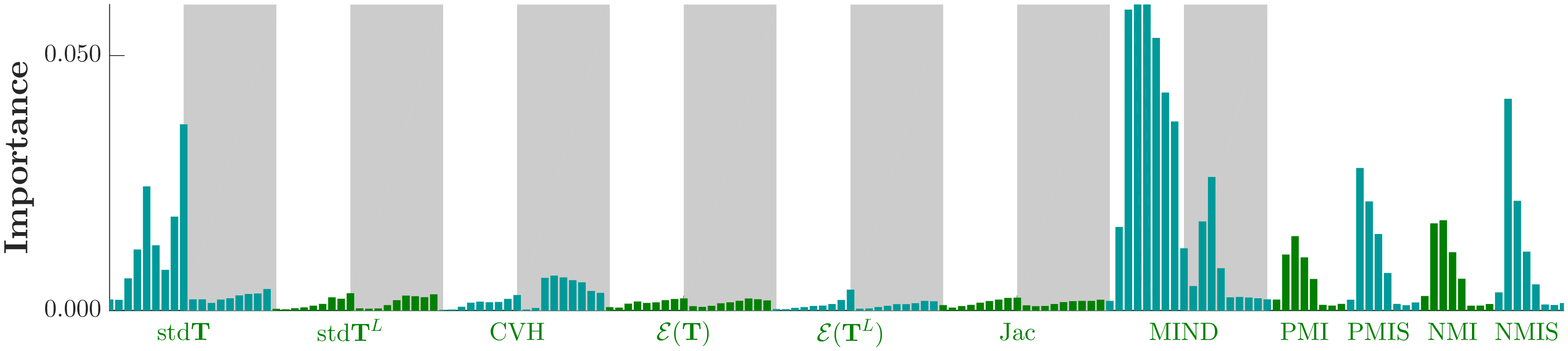}\label{fig:Imp:Sp100itr}}
	\subfigure[SPREAD 500 iterations]{\includegraphics[width=1\textwidth,trim={80 0 100 0cm},clip]{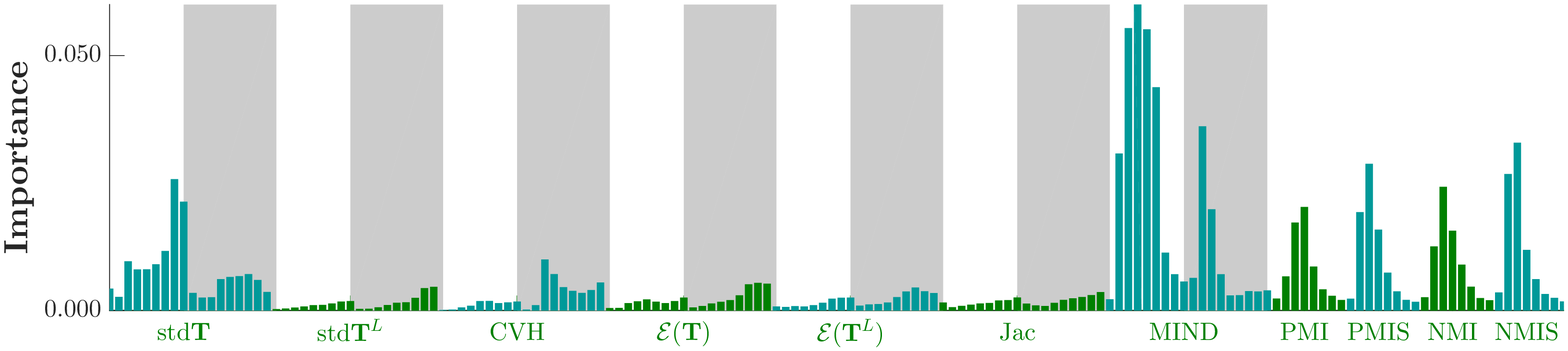}\label{fig:Imp:Sp500itr}}
	\subfigure[SPREAD 2000 iterations]{\includegraphics[width=1\textwidth,trim={80 0 100 0cm},clip]{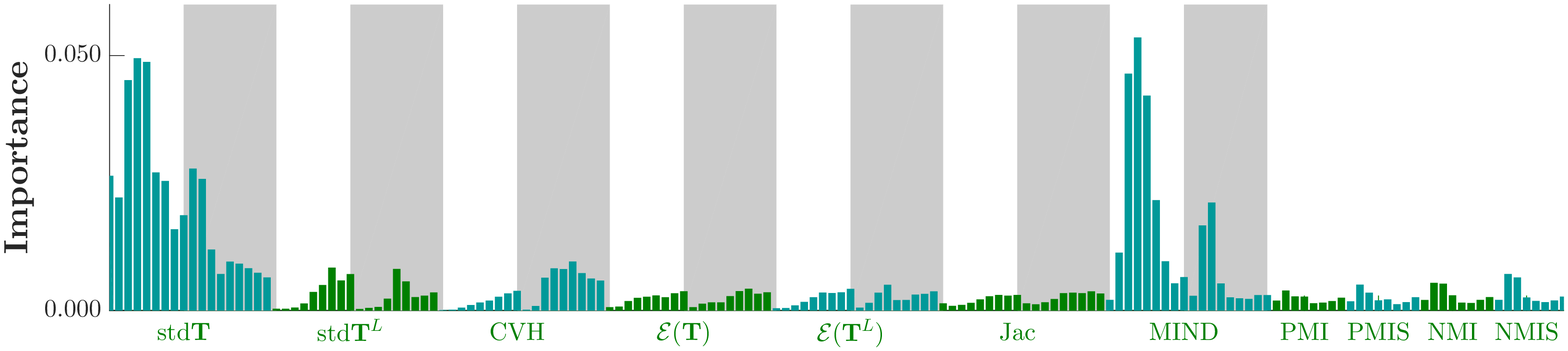}\label{fig:Imp:Sp2000itr}}

	\caption{Feature importance of the SPREAD combined experiment with different iterations. The contribution of all intensity features stay the same in all experiments, while some of the registration features contribute differently with respect to the number of iterations. White areas correspond to box averages, while shaded areas correspond to box maxima.}	
    \label{fig:ImportanceIterations}
\end{figure*}

\subsubsection{Excluding a single feature}
To further investigate the importance of the several features, we additionally perform an experiment where we leave one feature out of the combined feature set. The results are reported in Table \ref{tb:LOFO_SP}. In these experiments, feature redundancy can be found. For instance, MI has a large importance values in random forests, but if we leave that feature out, other  features can compensate for that. 

\begin{table*}[!tb]
	\centering
	\caption{Leave one feature out results of SPREAD data. The columns indicate the number of features ($N_f$), the mean absolute error (MAE), the accuracy (Acc) and the F1 score. The sub-indices c, p and w correspond to correct [0,3), poor [3,6) and wrong \mbox{[6, $\infty$) mm} classes, respectively.}
	\resizebox{\textwidth}{!}{
	\begin{tabular}{|L{\lengthMAE cm}|c|c|c|c|c|c|c|c|c|}
\hline
      &$N_f$   & MAE	& MAE\textsubscript{c}	& MAE\textsubscript{p}	& MAE\textsubscript{w}  & \hspace{.6mm}Acc\hspace{.6mm}  & F1\textsubscript{c} & F1\textsubscript{p} & F1\textsubscript{w}\\
  \hline
Combined & 158  &$1.07\pm1.86$  &$0.76\pm0.65$  &$1.47\pm1.22$  &$6.12\pm5.64$  &$90.7$  &$95.4$  &$38.1$  &$84.4$  \\
$  \hspace{0.3cm} -\mathrm{MIND}$ & 140  &$1.18\pm1.96$  &$0.83\pm0.66$  &$1.56\pm1.50$  &$6.70\pm5.69$  &$90.2$  &$95.1$  &$36.2$  &$83.0$  \\
$\hspace{0.3cm} -$MI & 126  &$1.10\pm1.98$  &$0.75\pm0.67$  &$1.54\pm1.30$  &$6.66\pm5.84$  &$90.6$  &$95.3$  &$37.0$  &$84.2$  \\
$\hspace{0.3cm} -\std\bm{T}$ & 140  &$1.08\pm1.86$  &$0.76\pm0.65$  &$1.46\pm1.18$  &$6.14\pm5.65$  &$90.7$  &$95.3$  &$38.1$  &$84.3$  \\
$\hspace{0.3cm} -\std\bm{T}^{\mathrm{L}}$ & 140  &$1.08\pm1.89$  &$0.76\pm0.65$  &$1.46\pm1.22$  &$6.21\pm5.73$  &$90.6$  &$95.3$  &$38.3$  &$83.7$  \\
$\hspace{0.3cm} -$CVH & 140  &$1.07\pm1.81$  &$0.75\pm0.65$  &$1.46\pm1.21$  &$6.06\pm5.98$  &$90.7$  &$95.4$  &$38.4$  &$84.3$  \\
$\hspace{0.3cm} -\mathcal{E}(\bm{T})$ & 140  &$1.07\pm1.86$  &$0.76\pm0.65$  &$1.46\pm1.21$  &$6.13\pm5.64$  &$90.7$  &$95.4$  &$38.2$  &$84.5$  \\
$\hspace{0.3cm} -\mathcal{E}(\bm{T}^{\mathrm{L}})$ & 140  &$1.08\pm1.85$  &$0.76\pm0.65$  &$1.47\pm1.22$  &$6.12\pm5.61$  &$90.6$  &$95.3$  &$37.5$  &$84.3$  \\
$\hspace{0.3cm} -$Jac & 140  &$1.08\pm1.87$  &$0.76\pm0.65$  &$1.49\pm1.31$  &$6.06\pm5.72$  &$90.7$  &$95.4$  &$37.9$  &$84.8$  \\
	\hline	
	\end{tabular}}
	\label{tb:LOFO_SP}	
\end{table*}

\subsubsection{Inter-database validation} \label{sec:inter-database}
To study the generalizability of the proposed system, instead of cross-validation on a single database, we perform training on the DIR-Lab-4DCT database and test it on the SPREAD database. As mentioned before, the SPREAD database consists of only inhale images but the DIR-Lab-4DCT database has images from inhale to exhale phases. Therefore, this makes the DIR-Lab-4DCT more suitable for training. The result of this experiment is available in Table \ref{tb:InterResult}. Once more, we can draw the conclusion that by combining both intensity and registration-based features, the regression performance can be improved. In contrast to the SPREAD experiment, this time it is observed that the registration features perform better than the intensity features.

To further evaluate the generalizability of the proposed method, we test it for different registration methods on a third independent test set, the DIR-Lab-COPDgene dataset. The regression forest is trained on a combination of the SPREAD and DIR-Lab-4DCT data. We evaluate two registration algorithms that achieved excellent performance in the EMPIRE10 challenge \citep{murphy2011evaluation}, i.e. the ANTs registration package \citep{avants2009advanced, tustisontwo} and \texttt{elastix} with advanced settings \citep{staring2010pulmonary}.

Prior to deformable registration we perform an affine registration using 5 resolutions and utilizing torso masks. For the deformable registration we use settings similar to the ones used in the EMPIRE10 challenge, specifically:

\textbf{ANTs-BSplineSyN:} With respect to the EMPIRE10 challenge we increased the number of iterations to 1000 for each of the 4 resolutions, using a 10\% sampling rate. This improved the performance on our data and considerably reduced the calculation time. As suggested in \cite{tustisontwo}, several preprocessing steps are used, including masking out the lungs, and inverting the image intensities and rescaling them between 0 and 1. Further settings include: registration model: symmetric diffeomorphic; dissimilarity metric: local cross correlation; number of resolutions: 4; maximum number of iterations: 1000; sampling: 10\% random samples; convergence threshold: 1e-6. The average TRE on DIR-Lab-COPDgene is \mbox{$1.90\pm2.86$ mm}.

\textbf{\texttt{elastix-advanced}:} Settings are adopted from \cite{staring2010pulmonary}. The most important ones are: registration model: B-spline; dissimilarity metric: normalized correlation; number of resolutions: 6; number of iterations: 1000; sampling: 2000 random samples;
B-spline grid spacing: [5, 5, 5] mm. The average TRE with this setting is \mbox{$3.39\pm4.30$ mm on the DIR-Lab-COPDgene dataset}.

Detailed parameter files for both registration methods are available via \href{http://elastix.bigr.nl/wiki/index.php/Parameter_file_database}{elastix.isi.uu.nl} (entry par0049) and \href{https://github.com/hsokooti/regun}{github.com/hsokooti/regun}. The calculation time of ANTs was about 60 hours per registration, comparing to 12 minutes for \texttt{elastix}.

In this experiment, the evaluation is performed only on
the landmarks locations, where Table \ref{tb:Composition} displays the distribution
of reference registration errors during testing. The results
of the experiments are given in Table \ref{tb:InterResult-Sp-COPD}. A scatter plot is also depicted in Fig. \ref{fig:ScatterPlot}. Similar to the previous inter-database experiment
(Table \ref{tb:InterResult}), the MAE and accuracy of the registration features are slightly better than the MAE and accuracy of the intensity-based features. However,
intensity features obtained better classification score in the wrong category. We conclude that the proposed method indeed generalizes to different settings of the same method (\texttt{elastix-advanced}), as well as registration methods with quite a different underlying transformation model (ANTs-BSplineSyN, which uses a symmetric diffeomorphic model).

\begin{table*}[!htb]
	\centering
	\caption{Regression results for the SPREAD data trained on the DIR-Lab-4DCT data with \texttt{elastix} using 20, 100, 500 and 2000 iterations. The columns indicate the number of features ($N_f$), the mean absolute error (MAE), the accuracy (Acc) and the F1 score. The sub-indices c, p and w correspond to correct [0,3), poor [3,6) and wrong \mbox{[6, $\infty$) mm} classes, respectively. 
	}
	\resizebox{\textwidth}{!}{
	\begin{tabular}{|L{\lengthMAE cm}|c|c|c|c|c|c|c|c|c|}
\hline
         &$N_f$ & MAE	&  MAE\textsubscript{c}	& MAE\textsubscript{p}	& MAE\textsubscript{w}  & \hspace{.6mm}Acc\hspace{.6mm}  & F1\textsubscript{c} & F1\textsubscript{p} & F1\textsubscript{w}\\
  \hline
Intensity & 50  &$1.90\pm3.63$  &$1.56\pm1.49$  &$1.26\pm1.01$  &$10.83\pm14.32$  &$71.0$  &$82.8$  &$21.7$  &$48.0$  \\
Registration & 108  &$1.62\pm3.59$  &$1.23\pm0.88$  &$1.13\pm0.81$  &$11.53\pm14.60$  &$77.1$  &$87.4$  &$27.7$  &$53.9$  \\
Combined & 158  &$1.73\pm3.56$  &$1.36\pm0.97$  &$1.14\pm0.83$  &$11.30\pm14.49$  &$77.2$  &$87.2$  &$26.0$  &$59.9$  \\
	\hline	
	\end{tabular}}
	\label{tb:InterResult}	
\end{table*}

\begin{table*}[!htb]
	\centering
	\caption{Regression results for the DIR-Lab-COPDgene data with \texttt{elastix-advanced} and ANTs-BSplineSyN registrations trained on the SPREAD and DIR-Lab-4DCT data. The columns indicate the number of features ($N_f$), the mean absolute error (MAE), the accuracy (Acc) and the F1 score. The sub-indices c, p and w correspond to correct [0,3), poor [3,6) and wrong \mbox{[6, $\infty$) mm} classes, respectively.}
	\resizebox{\textwidth}{!}{
	\begin{tabular}{|L{\lengthMAE cm}|c|c|c|c|c|c|c|c|c|}
\hline
         &$N_f$ & MAE	&  MAE\textsubscript{c}	& MAE\textsubscript{p}	& MAE\textsubscript{w}  & \hspace{.6mm}Acc\hspace{.6mm}  & F1\textsubscript{c} & F1\textsubscript{p} & F1\textsubscript{w}\\
  \hline

\texttt{elastix-advanced} &\phantom{x} &\phantom{x} &\phantom{x} &\phantom{x} &\phantom{x} &\phantom{x} &\phantom{x} &\phantom{x} &\phantom{x}\\
Intensity & 50        &$2.17\pm2.34$  &$1.69\pm1.35$  &$2.81\pm2.66$  &$5.15\pm4.44$  &$64.2$  &$77.9$  &$20.8$  &$64.6$  \\

Registration & 108       &$1.84\pm2.50$  &$1.31\pm1.66$  &$2.22\pm2.12$  &$5.36\pm4.29$  &$76.0$  &$87.6$  &$29.9$  &$57.6$  \\

Combined & 158        &$1.86\pm2.05$  &$1.50\pm1.16$  &$1.92\pm1.80$  &$4.48\pm4.21$  &$75.3$  &$86.9$  &$29.5$  &$66.1$  \\

\hline	

ANTs-BSplineSyN	&\phantom{x} &\phantom{x} &\phantom{x} &\phantom{x} &\phantom{x} &\phantom{x} &\phantom{x} &\phantom{x} &\phantom{x}\\
	
Intensity & 50    &$2.03\pm2.01$  &$1.80\pm1.25$  &$2.20\pm2.27$  &$5.30\pm5.38$  &$57.3$  &$71.6$  &$14.2$  &$62.7$  \\
  
Registration & 108    &$1.71\pm2.39$  &$1.43\pm1.98$  &$2.56\pm2.01$  &$5.06\pm4.67$  &$72.8$  &$85.5$  &$17.3$  &$38.5$  \\

Combined & 158    &$1.73\pm1.80$  &$1.52\pm1.23$  &$2.22\pm2.27$  &$4.45\pm4.40$  &$76.5$  &$87.3$  &$20.4$  &$59.7$  \\

\hline	

\end{tabular}}
\label{tb:InterResult-Sp-COPD}	
\end{table*}

\begin{figure}[!tb] 
	\centering	
	\subfigure{\includegraphics[width=1\columnwidth,trim={55 25 70 65},clip]{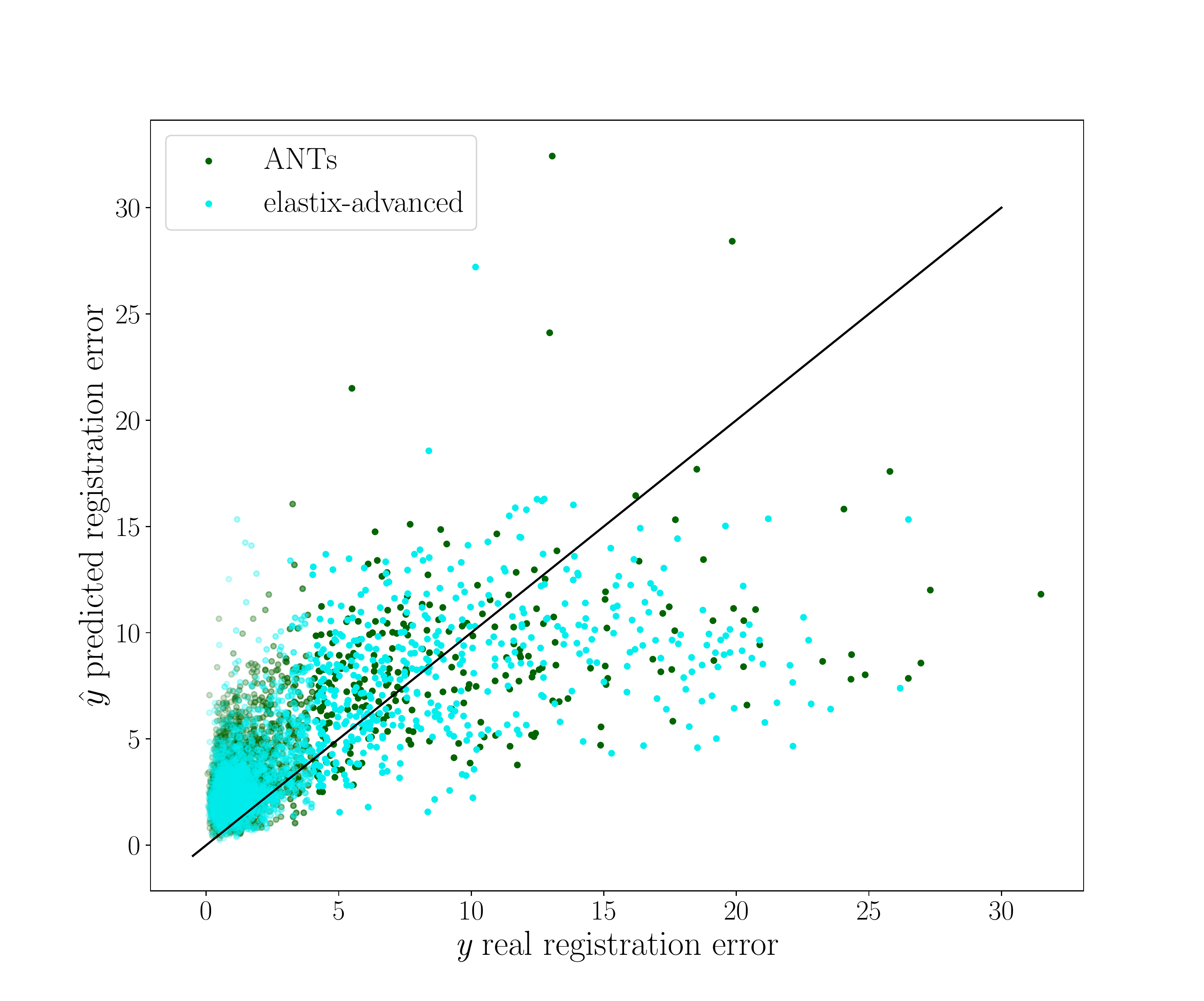}}
	\caption{Scatter plot of real and predicted registration errors in the DIR-Lab-COPDgene database using \texttt{elastix-advanced} and ANTs-BSplineSyN registration. In total, 3000 landmarks are shown for each registration.}
    \label{fig:ScatterPlot}
\end{figure}

\section{Discussion} \label{sec::discussion}
A system for quantitative error prediction of medical image registration is proposed and it is quantitatively evaluated on multiple chest CT datasets.
\subsection{Features} \label{sec::features}
In the intra-database (SPREAD) validation, it is observed that the single MIND feature can perform almost as good as the overall combined system. By adding MI and registration features, the results slightly improved. \citet{muenzing2012supervised} did not consider MIND in their feature set and found that the most important single features in their classification experiments are mutual information and Gaussian intensity, whereas, based on Table \ref{tb:SpreadFeatureResult} these features are less important than MIND in our experiments. Furthermore, the calculation time of MI for the whole image is about 3 h, as opposed to the calculation time of MIND, which is about \mbox{8 min} (\mbox{$\sim$3 min} MIND + \mbox{$\sim$5 min} pooling). Although less accurate, it is possible to reduce the calculation time
of the MI feature by calculating it over a single window and then aggregate by pooling.

The modality-dependent intensity features do not increase regression accuracy on the data used in our paper. Consequently more generally applicable modality-independent features can be used, even for mono-modal problems.

Table \ref{tb:SpreadResult}, \ref{tb:InterResult} and \ref{tb:InterResult-Sp-COPD} together suggests that features in the intensity and registration categories provide complementary information, and that a better system can be obtained in terms of MAE and accuracy by considering both intensity and registration-derived features. 

The intensity features were better predictors than the registration features in the intra-database experiment. However, in the inter-database experiment, the registration features outperform intensity features in terms of total accuracy and MAE. The same observation can be made for the average F1 score in the inter-database experiments using \texttt{elastix} (See Table \ref{tb:InterResult}, \ref{tb:InterResult-Sp-COPD}). For ANTs (Table \ref{tb:InterResult-Sp-COPD}), the average F1 score of the intensity-based features was slightly higher than that of the registration-based features.

The registration features contribute differently with respect to the number of iterations (See Fig. \ref{fig:ImportanceIterations}). The features $\std\bm{T}^{\mathrm{L}}$ and $\mathcal{E}(\bm{T}^{\mathrm{L}})$ play important roles when the number of iterations is not enough for convergence. When the number of iterations increases, the contribution of $\std\bm{T}$ and CVH go up. In the work of \citet{muenzing2012supervised}, only one registration feature, Jac, was used and they reported that the impact of this feature is relatively low in comparison with intensity-based features. We observed the same result for Jac, but it should be pointed out that the range of Jac in our database was [0.3, 3.9] so voxels with negative or very large values were not encountered.

Feature pooling improves the regression results in all evaluation measures, due to the addition of contextual information. In some features like $\std\bm{T}$, average pooling contributed more to the regression performance, while in features like CVH, maximum pooling had a higher importance value (See Fig. \ref{fig:Imp:Sp2000itr}).

As can be seen in Table \ref{tb:LOFO_SP}, the proposed combined system has redundant features. Hence, by removing a single feature, the system is still able to predict the registration error with almost equal MAE as the total system. However, removing these features may decrease the generalizability of the system. For example, looking at the feature $\mathcal{E}(\bm{T}^{\mathrm{L}})$ in Fig. \ref{fig:Importance} we see that its contribution is relatively small overall. However, in Fig. \ref{fig:ImportanceIterations} it can be seen that while it is less important for better registration results (100, 500 and 2000 iterations), it is still important for poor registration results (20 iterations).

Considering the results in both intra and inter-database experiments (Table \ref{tb:SpreadResult}, \ref{tb:InterResult} and \ref{tb:InterResult-Sp-COPD}), the conclusion to be drawn is that the proposed combined feature sets is general and robust.

\subsection{Quantitative validation }
Commonly, in image registration tasks, the distribution of registration errors is not balanced as can be seen in Table \ref{tb:CompositionTraining}. 

In the SPREAD experiment, Table \ref{tb:SpreadResult} reports that in the combined experiment, the MAE of the correct and poor classes are $0.76 \pm 0.65 $ and $1.47 \pm 1.22 $, respectively. On the contrary, the MAE of the wrong class is $6.12 \pm 5.64 $. It is expected that the regression error of values of the wrong class is relatively larger than that of the other classes. However,
it should be emphasized that only $3.9\%$ of samples are available to make a regression model between 6 and \mbox{81.8 mm}. We tried to add more samples and make the distribution more balanced by performing registrations with different number of iterations, but there is still room for improvement for the wrong class by adding more samples and data.

In terms of classification, we obtained F1 scores of $95.4\%$, $38.1\%$ and $84.4\%$ in the classes correct, poor and wrong, respectively (Table \ref{tb:SpreadResult}). For the wrong class, which is arguably most important for clinical application, the precision and recall are $84.6\%$ and $84.3\%$, respectively. This means that $84.6\%$ of all samples predicted to be over 6 mm are correct and the proposed method caught $84.3\%$ of larger registration errors. \citet{muenzing2012supervised} obtained F1 scores of  $95.3\%$, $73.8\%$ and $86.6\%$ in the classes [0,2), [2,5) and \mbox{[5, $\infty$) mm}. They achieved a better F1 score in the poor class and they also reported zero overlap between the correct and wrong classes. However, the comparison between the two methods is not easy because of the differences in the data. For example, the slice thickness in SPREAD is \mbox{2.5 mm}, while it is \mbox{1 mm} for Muenzing's data, which may affect the performance especially in the poor class. Moreover, we generated the classes by thresholding the regression values. Thus, the forests are optimized for regression not for classification.

\subsection{Qualitative validation }
\citet{muenzing2014dirboost} generated an uncertainty map by spatial interpolation of landmark-based quality estimates. On the contrary, our proposed system, which is trained on landmark locations, can be applied in all regions of the image. We showed this for two example images, see Fig. \ref{fig:GT}. It can be easily visualized that in the blue region, images are matched correctly. On the other hand, by tracking the vessels in the red region misalignment can be seen. Another note about the prediction is that there are no abrupt changes, and error varies smoothly from blue to yellow and then red, even though the error is predicted for each voxel independently.

Another example is given in Fig.  \ref{fig:Qualitative}(a-d). Although all landmarks indicate that the registration error is small in this slice, the quantitative results found several misregistered regions, which implies that few landmarks may not be sufficient to assess the registration quality of the whole image. In Fig. \ref{fig:Qualitative}(e, f), it can be observed that the performance in the homogeneous area (left side of the images) is as good as the performance in the area with structure. The main reason of acceptable performance in the homogeneous area is that the training samples consist of landmarks as well as their neighborhood region, which can be homogeneous. Thus, the system is trained both for homogeneous regions and regions with structure.

Another example is given in Fig \ref{fig:Qualitative}(g, h), where the proposed system is not able to predict the registration error correctly because of a shift in the slice direction.

\begin{figure}[!tb] 
	\centering
	
	\subfigure {\includegraphics[width=.998\columnwidth]{ColorBar.eps}}
	{\vspace{-6mm}} 
	\addtocounter{subfigure}{-1}
	
	\subfigure[Sample 1: ${I}_F$]{\includegraphics[width=.49\columnwidth]	{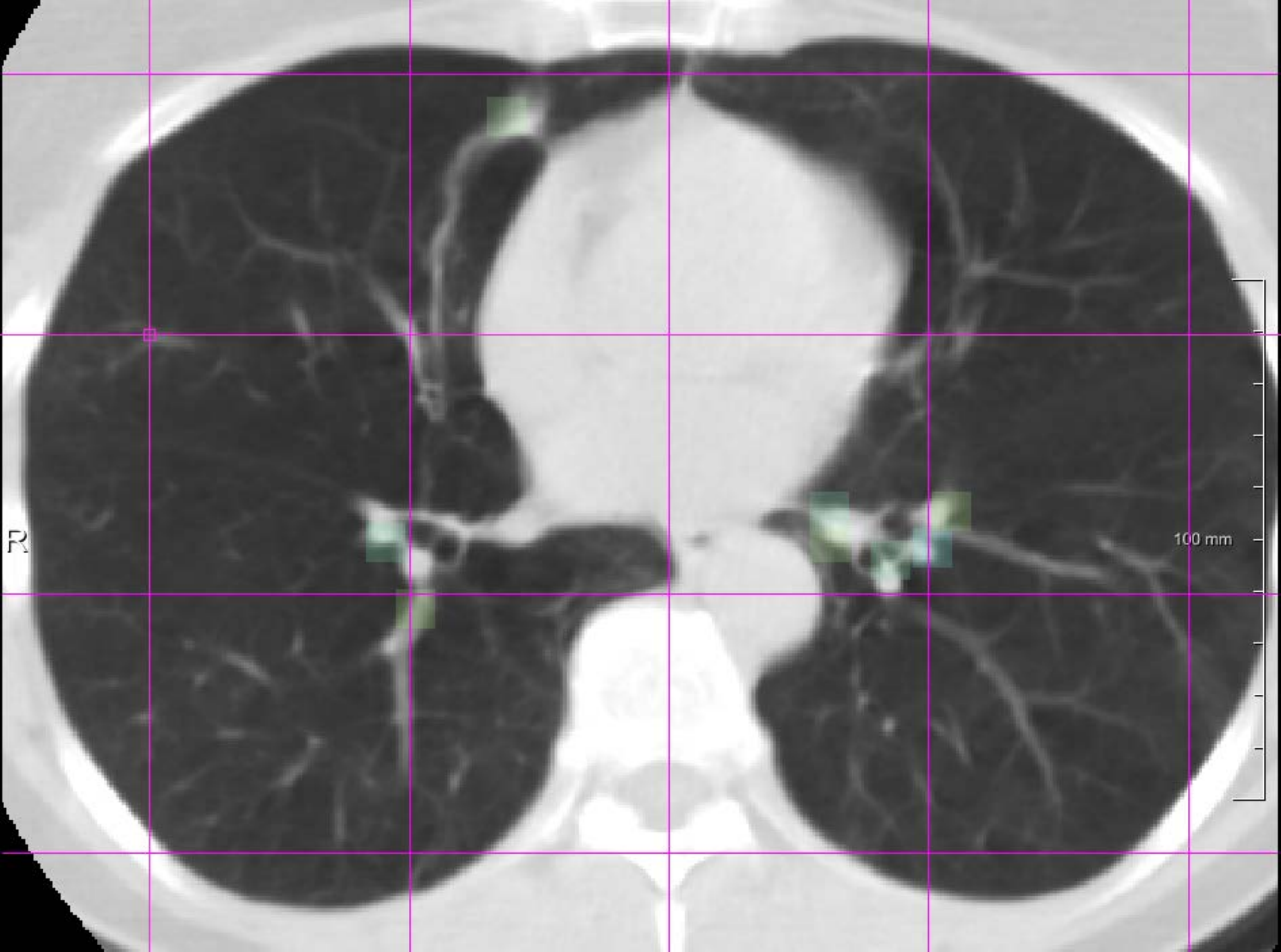}}
	\subfigure[Sample 1: ${I}_M(\bf{T^\mathrm{b}})$]{\includegraphics[width=.49\columnwidth]	{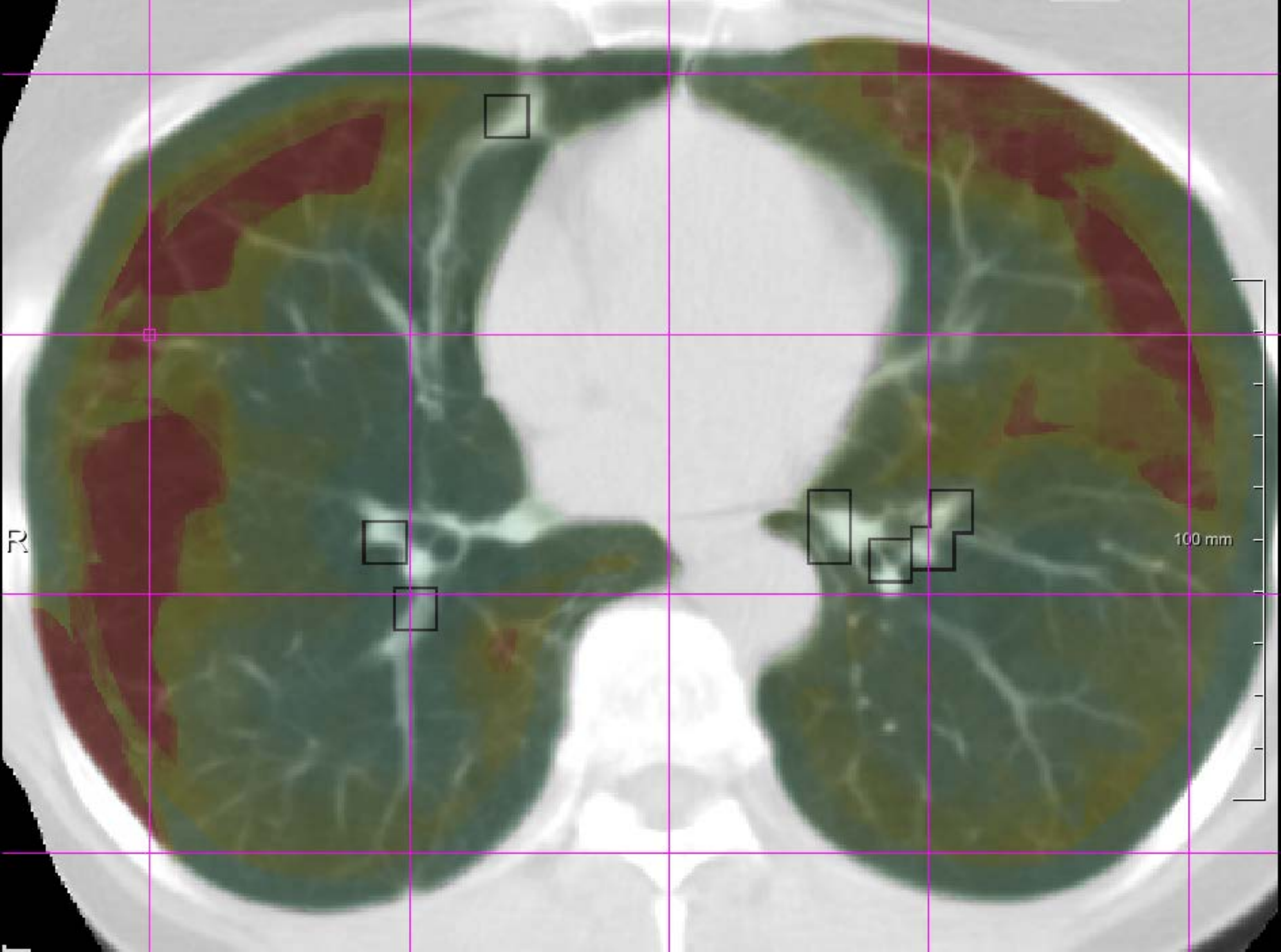}}		
	\subfigure[Magnification of (a)]{\includegraphics[width=.49\columnwidth]	{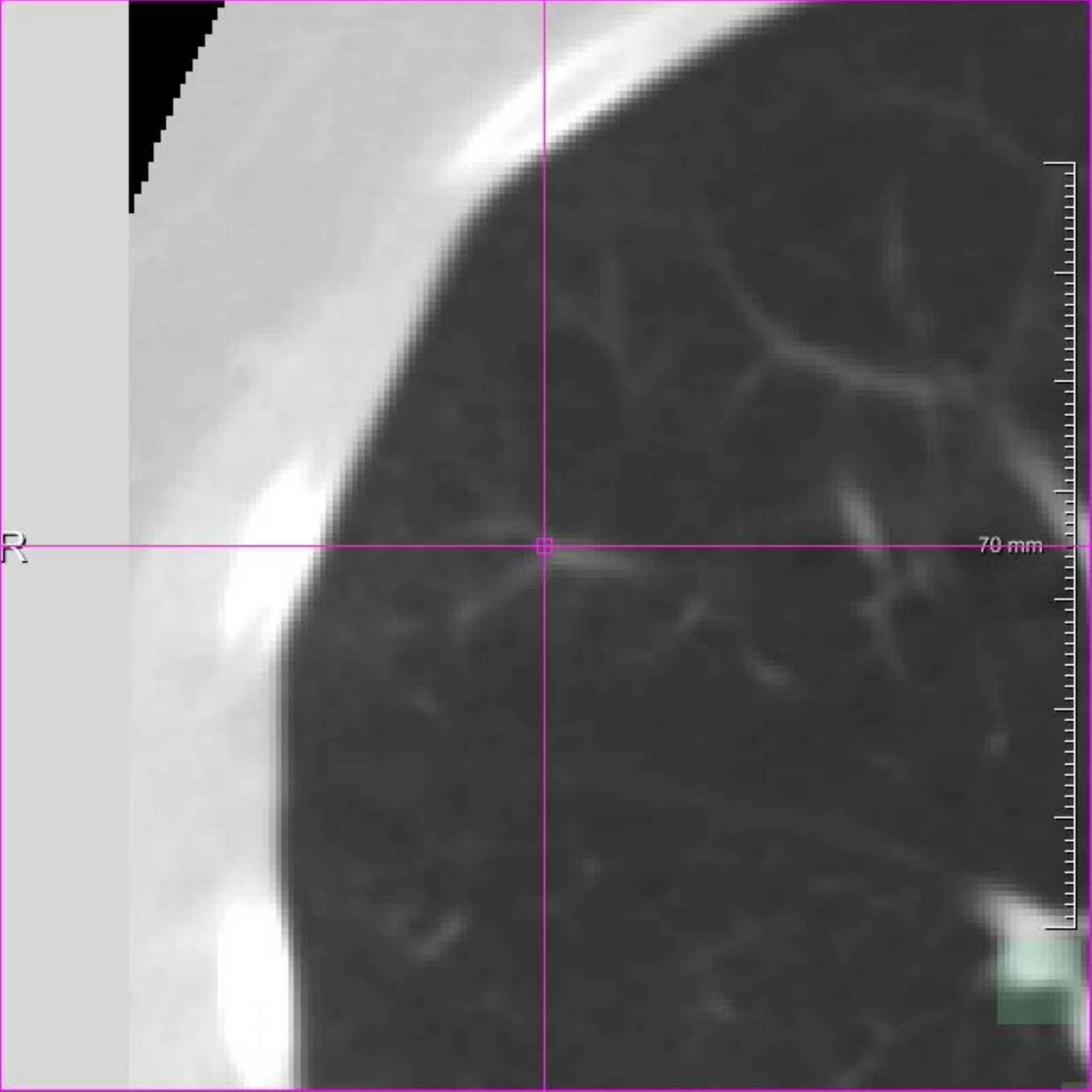}}
	\subfigure[Magnification of (b)]{\includegraphics[width=.49\columnwidth]	{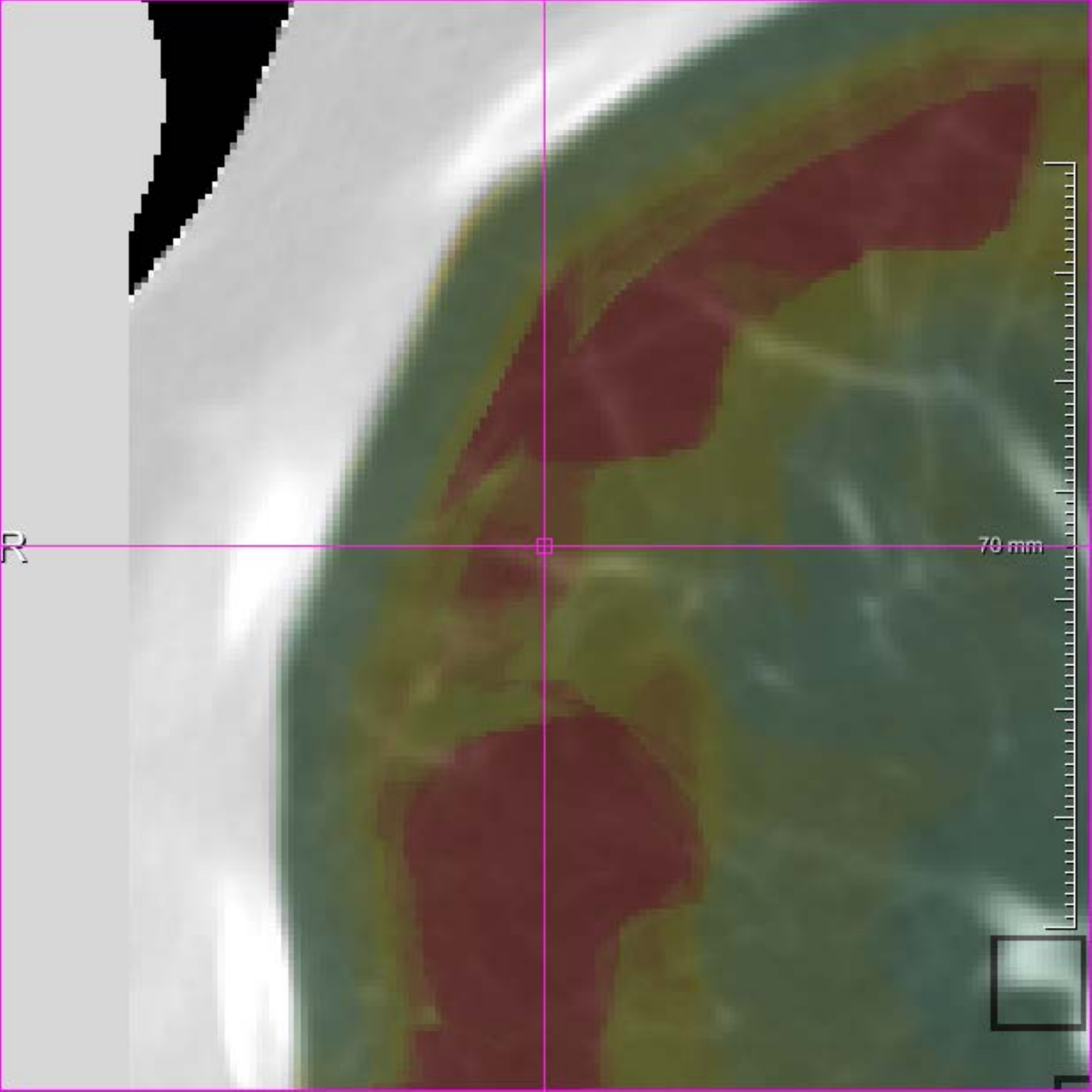}}

	\subfigure[Sample 2: ${I}_F$]{\includegraphics[width=.49\columnwidth]	{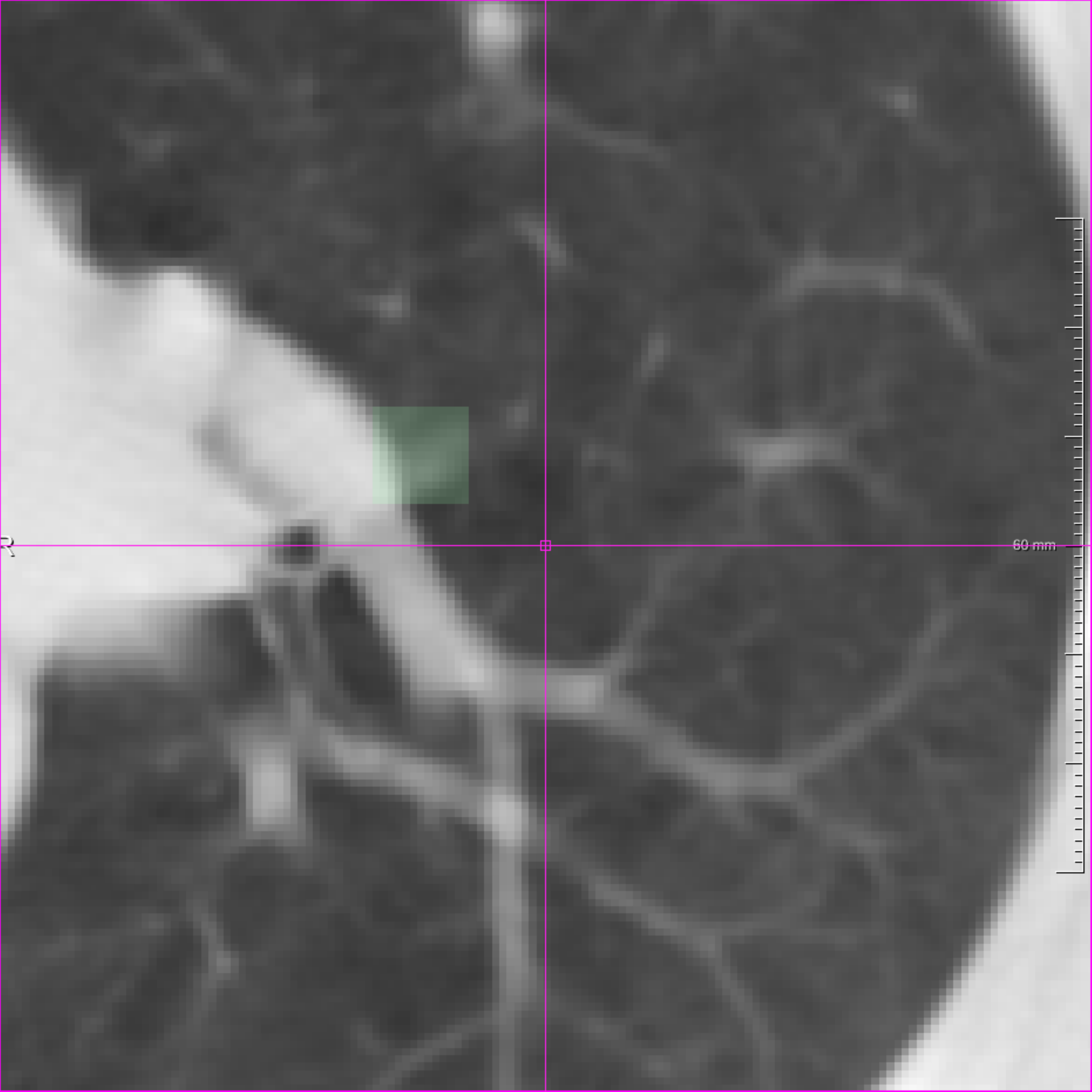}}
	\subfigure[Sample 2: ${I}_M(\bf{T^\mathrm{b}})$]{\includegraphics[width=.49\columnwidth]	{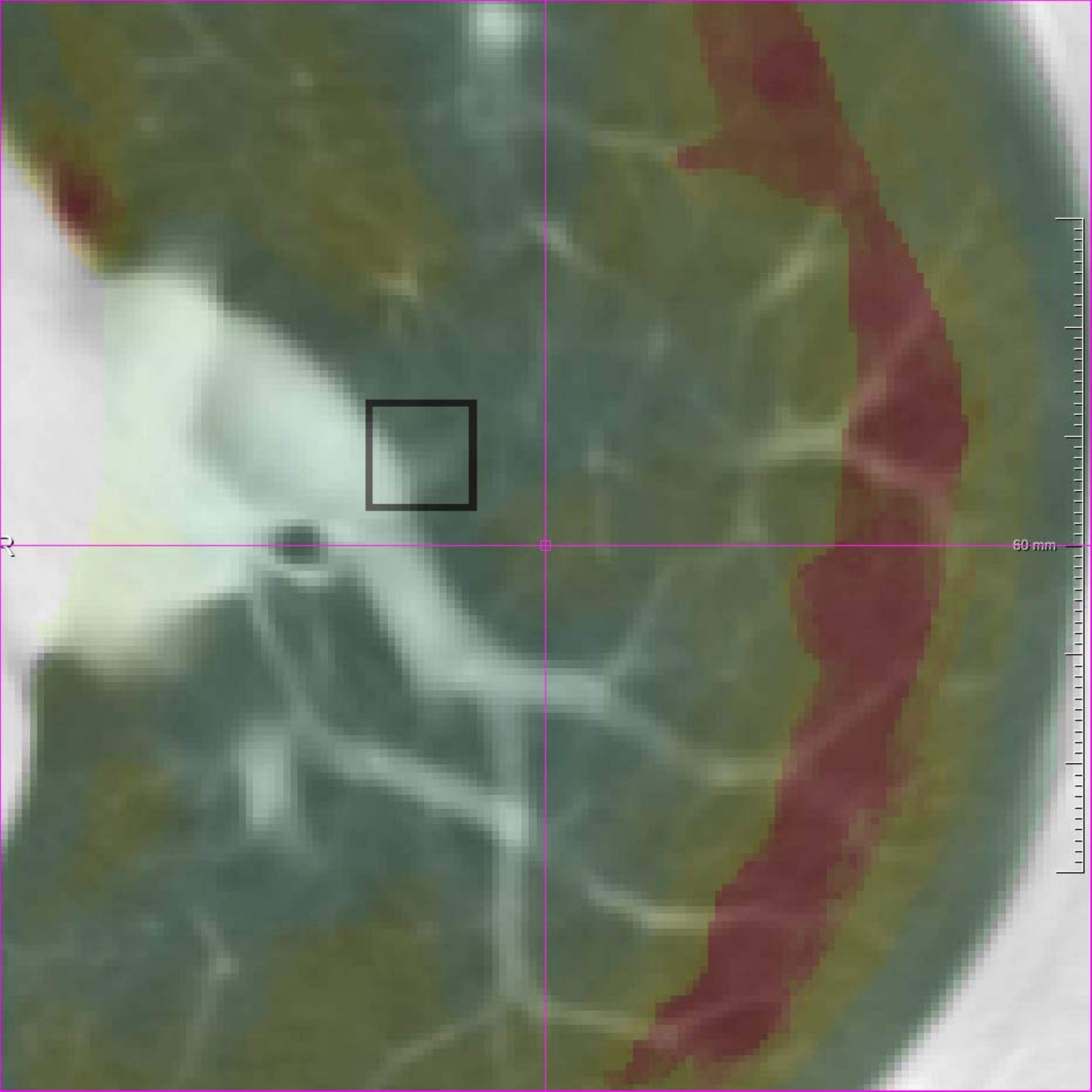}}

		\subfigure[Sample 3: ${I}_F$]{\includegraphics[width=.49\columnwidth]	{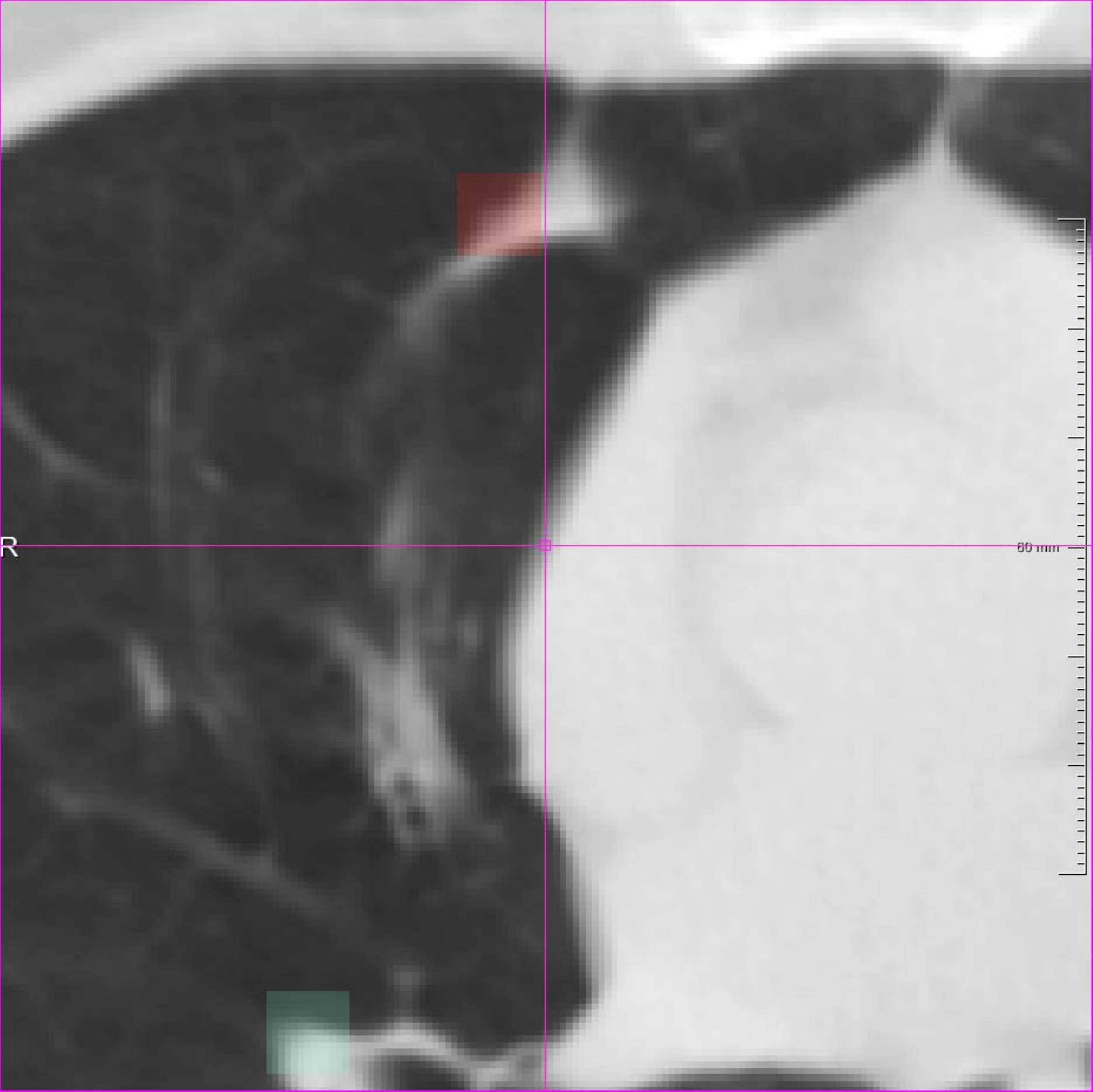}}
	\subfigure[Sample 3: ${I}_M(\bf{T^\mathrm{b}})$]{\includegraphics[width=.49\columnwidth]	{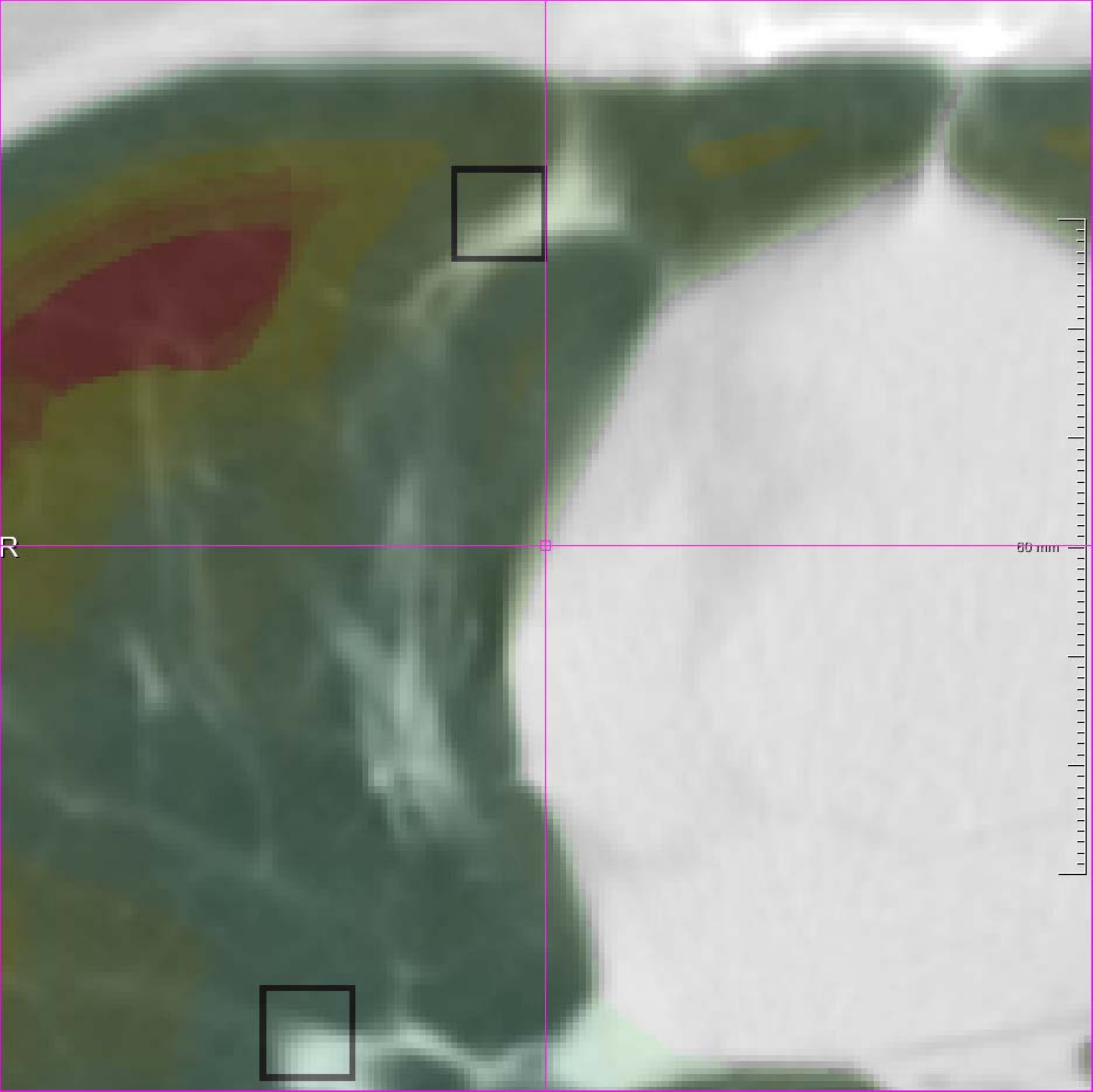}}	

	\caption{Several samples from the SPREAD dataset. The left column shows the fixed image with the ground truth registration error overlaid in color. The right column shows the moving image after registration with the registration error predicted by the proposed method overlaid in color.}	
\label{fig:Qualitative}
\end{figure}

\subsection{Limitations}

\textbf{Discrete optimization:} If the optimization method is less or not dependent on the initial state, for instance for discrete optimization methods \citep{glocker2008optical, heinrich2016deformable}, many of the proposed registration features, which are generated by varying the initial transformation of the registration, are not informative anymore. In such cases, instead of $\std{\bm{T}}$ or $\std{\bm{T}^\mathrm{L}}$, other measures can be used. For example, by utilizing the adaptive mean-shift algorithm, the local standard deviation of the displacement distribution can be calculated \citep{heinrich2016deformable}.

\textbf{Anatomical changes:} The proposed method is trained in such a way that any dissimilarity between the fixed and moving images is counted as misalignment in registration. In case of anatomical changes this assumption may be invalid, but typically prior knowledge of the underlying anatomy is required to determine which regions are allowed to be "misaligned" because of anatomical changes and which are not \citep{muenzing2018larvalign}. The proposed method highlights all changes, coming from misalignment or from anatomical change.

\subsection{Future work}
In the proposed method we predict the misalignment as an Euclidean distance in millimeters, rather than a 3D vector representing residual displacement. This is mostly because the features used in the system are not direction-wise, especially the local intensity features. The use of features that include directional information may help the system to be used in predicting the registration error in each direction, which is then effectively a new registration method.

The proposed method was tested on chest CT scans. Since the proposed features are generic and modality-independent, the overall method can in principle be applied to other modality data from other anatomical regions. The performance in such cases however remains to be investigated.

The uncertainty of affine registration is not measured in this work. Defining a gold standard for this mid-phase result is a complex task. However, extending the experiments to other databases where only affine transformations are applicable can be done in the future.

Instead of manually defined features, it is possible to use convolutional neural networks, which can learn features automatically. \citet{eppenhof2017supervised} predicted the Euclidean distance of registration error. Our own work on CNNs for registration \citep{sokooti2017nonrigid} can also be modified to predict registration uncertainty in a direction-wise manner. Both methods are trained only based on intensity, where the current paper shows that registration-derived information still contributes to a better regression. Thus, adding registration information to the neural networks should probably be considered as well.

A larger set of corresponding points annotated more densely throughout the scan could potentially also benefit training of the regression forest. In addition, experimenting on multi-modality data and investigating the contribution of all introduced features on them are future plans of this work.

Finally, the uncertainty map produced by the proposed method may be exploited to improve local registration results.

\section{Conclusion}

In this paper we proposed a method based on random regression forests to predict registration accuracy on chest CT scans from registration-based as well as intensity-based features. We introduced the variation in registration result from differences in initialization ($\std{\bm{T}}$) and CVH, which showed high feature importance in several experiments.  Registration-based features provided additional information on registration error with respect to intensity-based features.

The regression method was evaluated on data from the SPREAD study and predicted the registration error with a mean absolute error of 1.07 $\pm$ 1.86 mm. The proposed method gained promising results on inter-database validation with a regression error of \mbox{1.76 $\pm$ 2.59 mm}. 

\section*{Acknowledgments}
This work is financed by the Netherlands Organization for Scientific
Research (NWO), project 13351. Ben Glocker received funding from the European Research Council (ERC) under the European Union’s Horizon 2020 research and innovation programme (grant agreement no. 757173, project MIRA, ERC-2017-STG).
Dr. M.E. Bakker and J. Stolk are acknowledged for providing a ground truth for the SPREAD study data used in this paper. We would like to thank Dr. R. Castillo and T. Guerrero for providing the DIR-Lab database.

\section{References}
\bibliographystyle{elsarticle-harv}

\bibliography{Uncertainty}

\begin{thebibliography}{47}
\expandafter\ifx\csname natexlab\endcsname\relax\def\natexlab#1{#1}\fi
\expandafter\ifx\csname url\endcsname\relax
  \def\url#1{\texttt{#1}}\fi
\expandafter\ifx\csname urlprefix\endcsname\relax\def\urlprefix{URL }\fi

\bibitem[{Avants et~al.(2009)Avants, Tustison, and Song}]{avants2009advanced}
Avants, B.~B., Tustison, N., Song, G., 2009. Advanced normalization tools
  (ants). Insight J. 2, 1--35.

\bibitem[{Breiman(2001)}]{breiman2001random}
Breiman, L., 2001. Random forests. Machine Learning 45~(1), 5--32.

\bibitem[{Castillo et~al.(2013)Castillo, Castillo, Fuentes, Ahmad, Wood,
  Ludwig, and Guerrero}]{castillo2013reference}
Castillo, R., Castillo, E., Fuentes, D., Ahmad, M., Wood, A.~M., Ludwig, M.~S.,
  Guerrero, T., 2013. A reference dataset for deformable image registration
  spatial accuracy evaluation using the copdgene study archive. Physics in
  Medicine \& Biology 58~(9), 2861.

\bibitem[{Castillo et~al.(2009)Castillo, Castillo, Guerra, Johnson, McPhail,
  Garg, and Guerrero}]{castillo2009framework}
Castillo, R., Castillo, E., Guerra, R., Johnson, V.~E., McPhail, T., Garg,
  A.~K., Guerrero, T., 2009. A framework for evaluation of deformable image
  registration spatial accuracy using large landmark point sets. Physics in
  Medicine and Biology 54~(7), 1849.

\bibitem[{Datteri and Dawant(2012)}]{datteri2012automatic}
Datteri, R.~D., Dawant, B.~M., 2012. Automatic detection of the magnitude and
  spatial location of error in non-rigid registration. In: Biomedical Image
  Registration. Springer, pp. 21--30.

\bibitem[{Eppenhof and Pluim(2017)}]{eppenhof2017supervised}
Eppenhof, K.~A., Pluim, J.~P., 2017. Supervised local error estimation for
  nonlinear image registration using convolutional neural networks. In: SPIE
  Medical Imaging. International Society for Optics and Photonics, pp.
  101331U--101331U.

\bibitem[{Forsberg et~al.(2011)Forsberg, Rathi, Bouix, Wassermann, Knutsson,
  and Westin}]{forsberg2011improving}
Forsberg, D., Rathi, Y., Bouix, S., Wassermann, D., Knutsson, H., Westin,
  C.-F., 2011. Improving registration using multi-channel diffeomorphic demons
  combined with certainty maps. In: Multimodal Brain Image Analysis. Springer,
  pp. 19--26.

\bibitem[{Gass et~al.(2015)Gass, Szekely, and Goksel}]{gass2015consistency}
Gass, T., Szekely, G., Goksel, O., 2015. Consistency-based rectification of
  nonrigid registrations. Journal of Medical Imaging 2~(1), 014005--014005.

\bibitem[{Glocker et~al.(2008)Glocker, Paragios, Komodakis, Tziritas, and
  Navab}]{glocker2008optical}
Glocker, B., Paragios, N., Komodakis, N., Tziritas, G., Navab, N., 2008.
  Optical flow estimation with uncertainties through dynamic mrfs. In: Computer
  Vision and Pattern Recognition, 2008. CVPR 2008. IEEE Conference on. IEEE,
  pp. 1--8.

\bibitem[{Glocker et~al.(2014)Glocker, Zikic, and Haynor}]{glocker2014robust}
Glocker, B., Zikic, D., Haynor, D.~R., 2014. Robust registration of
  longitudinal spine ct. In: International Conference on Medical Image
  Computing and Computer-Assisted Intervention. Springer, pp. 251--258.

\bibitem[{Gunay et~al.(2017)Gunay, Luu, Moelker, Walsum, and
  Klein}]{gunay2017semi}
Gunay, G., Luu, M.~H., Moelker, A., Walsum, T., Klein, S., 2017. Semi-automated
  registration of pre-and intra-operative {CT} for image-guided percutaneous
  liver tumor ablation interventions. Medical Physics.

\bibitem[{Heinrich et~al.(2012)Heinrich, Jenkinson, Bhushan, Matin, Gleeson,
  Brady, and Schnabel}]{heinrich2012mind}
Heinrich, M.~P., Jenkinson, M., Bhushan, M., Matin, T., Gleeson, F.~V., Brady,
  M., Schnabel, J.~A., 2012. {MIND}: Modality independent neighbourhood
  descriptor for multi-modal deformable registration. Medical Image Analysis
  16~(7), 1423--1435.

\bibitem[{Heinrich et~al.(2016)Heinrich, Simpson, Papie{\.z}, Brady, and
  Schnabel}]{heinrich2016deformable}
Heinrich, M.~P., Simpson, I.~J., Papie{\.z}, B.~W., Brady, M., Schnabel, J.~A.,
  2016. Deformable image registration by combining uncertainty estimates from
  supervoxel belief propagation. Medical Image Analysis 27, 57--71.

\bibitem[{Hub and Karger(2013)}]{hub2013estimation}
Hub, M., Karger, C., 2013. Estimation of the uncertainty of elastic image
  registration with the {D}emons algorithm. Physics in Medicine and Biology
  58~(9), 3023.

\bibitem[{Hub et~al.(2009)Hub, Kessler, and Karger}]{hub2009stochastic}
Hub, M., Kessler, M.~L., Karger, C.~P., 2009. A stochastic approach to estimate
  the uncertainty involved in {B}-spline image registration. IEEE Transactions
  on Medical Imaging 28~(11), 1708--1716.

\bibitem[{Klein et~al.(2009)Klein, Pluim, Staring, and
  Viergever}]{klein2009adaptive}
Klein, S., Pluim, J.~P., Staring, M., Viergever, M.~A., 2009. Adaptive
  stochastic gradient descent optimisation for image registration.
  International Journal of Computer Vision 81~(3), 227--239.

\bibitem[{Klein et~al.(2010)Klein, Staring, Murphy, Viergever, and
  Pluim}]{klein2010elastix}
Klein, S., Staring, M., Murphy, K., Viergever, M.~A., Pluim, J.~P., 2010.
  Elastix: A toolbox for intensity-based medical image registration. IEEE
  Transactions on Medical Imaging 29~(1), 196--205.

\bibitem[{Kybic(2010)}]{kybic2010bootstrap}
Kybic, J., 2010. Bootstrap resampling for image registration uncertainty
  estimation without ground truth. IEEE Transactions on Image Processing
  19~(1), 64--73.

\bibitem[{Liaw et~al.(2002)Liaw, Wiener, et~al.}]{liaw2002classification}
Liaw, A., Wiener, M., et~al., 2002. Classification and regression by
  randomforest. R news 2~(3), 18--22.

\bibitem[{Lotfi et~al.(2013)Lotfi, Tang, Andrews, and
  Hamarneh}]{lotfi2013improving}
Lotfi, T., Tang, L., Andrews, S., Hamarneh, G., 2013. Improving probabilistic
  image registration via reinforcement learning and uncertainty evaluation. In:
  Machine Learning in Medical Imaging. Springer, pp. 187--194.

\bibitem[{Luo et~al.(2017)Luo, Popuri, Cobzas, Ding, Wells, and
  Sugiyama}]{luo2017misdirected}
Luo, J., Popuri, K., Cobzas, D., Ding, H., Wells, W.~M., Sugiyama, M., 2017.
  Misdirected registration uncertainty. arXiv preprint arXiv:1704.08121.

\bibitem[{Muenzing et~al.(2018)Muenzing, Strauch, Truman, B{\"u}hler, Thum, and
  Merhof}]{muenzing2018larvalign}
Muenzing, S.~E., Strauch, M., Truman, J.~W., B{\"u}hler, K., Thum, A.~S.,
  Merhof, D., 2018. Larvalign: Aligning gene expression patterns from the
  larval brain of drosophila melanogaster. Neuroinformatics 16~(1), 65--80.

\bibitem[{Muenzing et~al.(2012)Muenzing, van Ginneken, Murphy, and
  Pluim}]{muenzing2012supervised}
Muenzing, S.~E., van Ginneken, B., Murphy, K., Pluim, J.~P., 2012. Supervised
  quality assessment of medical image registration: Application to
  intra-patient {CT} lung registration. Medical Image Analysis 16~(8),
  1521--1531.

\bibitem[{Muenzing et~al.(2014)Muenzing, van Ginneken, Viergever, and
  Pluim}]{muenzing2014dirboost}
Muenzing, S.~E., van Ginneken, B., Viergever, M.~A., Pluim, J.~P., 2014.
  Dirboost--an algorithm for boosting deformable image registration:
  Application to lung {CT} intra-subject registration. Medical Image Analysis
  18~(3), 449--459.

\bibitem[{Murphy et~al.(2011{\natexlab{a}})Murphy, van Ginneken, Klein,
  Staring, de~Hoop, Viergever, and Pluim}]{murphy2011semi}
Murphy, K., van Ginneken, B., Klein, S., Staring, M., de~Hoop, B.~J.,
  Viergever, M.~A., Pluim, J.~P., 2011{\natexlab{a}}. Semi-automatic
  construction of reference standards for evaluation of image registration.
  Medical Image Analysis 15~(1), 71--84.

\bibitem[{Murphy et~al.(2011{\natexlab{b}})Murphy, Van~Ginneken, Reinhardt,
  Kabus, Ding, Deng, Cao, Du, Christensen, Garcia,
  et~al.}]{murphy2011evaluation}
Murphy, K., Van~Ginneken, B., Reinhardt, J.~M., Kabus, S., Ding, K., Deng, X.,
  Cao, K., Du, K., Christensen, G.~E., Garcia, V., et~al., 2011{\natexlab{b}}.
  Evaluation of registration methods on thoracic {CT}: the {EMPIRE10}
  challenge. IEEE Transactions on Medical Imaging 30~(11), 1901--1920.

\bibitem[{Murphy et~al.(2012)Murphy, Salguero, Siebers, Staub, and
  Vaman}]{murphy2012method}
Murphy, M.~J., Salguero, F.~J., Siebers, J.~V., Staub, D., Vaman, C., 2012. A
  method to estimate the effect of deformable image registration uncertainties
  on daily dose mapping. Medical Physics 39~(2), 573--580.

\bibitem[{Park et~al.(2004)Park, Bland, Brock, and Meyer}]{park2004adaptive}
Park, H., Bland, P.~H., Brock, K.~K., Meyer, C.~R., 2004. Adaptive registration
  using local information measures. Medical Image Analysis 8~(4), 465--473.

\bibitem[{Pedregosa et~al.(2011)Pedregosa, Varoquaux, Gramfort, Michel,
  Thirion, Grisel, Blondel, Prettenhofer, Weiss, Dubourg, Vanderplas, Passos,
  Cournapeau, Brucher, Perrot, and Duchesnay}]{scikit-learn}
Pedregosa, F., Varoquaux, G., Gramfort, A., Michel, V., Thirion, B., Grisel,
  O., Blondel, M., Prettenhofer, P., Weiss, R., Dubourg, V., Vanderplas, J.,
  Passos, A., Cournapeau, D., Brucher, M., Perrot, M., Duchesnay, E., 2011.
  Scikit-learn: Machine learning in {P}ython. Journal of Machine Learning
  Research 12, 2825--2830.

\bibitem[{Risholm et~al.(2013)Risholm, Janoos, Norton, Golby, and
  Wells}]{risholm2013bayesian}
Risholm, P., Janoos, F., Norton, I., Golby, A.~J., Wells, W.~M., 2013. Bayesian
  characterization of uncertainty in intra-subject non-rigid registration.
  Medical Image Analysis 17~(5), 538--555.

\bibitem[{Rohde et~al.(2003)Rohde, Aldroubi, and Dawant}]{rohde2003adaptive}
Rohde, G.~K., Aldroubi, A., Dawant, B.~M., 2003. The adaptive bases algorithm
  for intensity-based nonrigid image registration. IEEE Transactions on Medical
  Imaging 22~(11), 1470--1479.

\bibitem[{Rueckert et~al.(1999)Rueckert, Sonoda, Hayes, Hill, Leach, and
  Hawkes}]{rueckert1999nonrigid}
Rueckert, D., Sonoda, L.~I., Hayes, C., Hill, D.~L., Leach, M.~O., Hawkes,
  D.~J., 1999. Nonrigid registration using free-form deformations: Application
  to breast {MR} images. IEEE Transactions on Medical Imaging 18~(8), 712--721.

\bibitem[{Saygili et~al.(2016)Saygili, Staring, and
  Hendriks}]{saygili2016confidence}
Saygili, G., Staring, M., Hendriks, E.~A., 2016. Confidence estimation for
  medical image registration based on stereo confidences. IEEE Transactions on
  Medical Imaging 35~(2), 539--549.

\bibitem[{Schlachter et~al.(2016)Schlachter, Fechter, Jurisic, Schimek-Jasch,
  Oehlke, Adebahr, Birkfellner, Nestle, and
  B{\"u}hler}]{schlachter2016visualization}
Schlachter, M., Fechter, T., Jurisic, M., Schimek-Jasch, T., Oehlke, O.,
  Adebahr, S., Birkfellner, W., Nestle, U., B{\"u}hler, K., 2016. Visualization
  of deformable image registration quality using local image dissimilarity.
  IEEE Transactions on Medical Imaging 35~(10), 2319--2328.

\bibitem[{Schnabel et~al.(2001)Schnabel, Rueckert, Quist, Blackall,
  Castellano-Smith, Hartkens, Penney, Hall, Liu, Truwit,
  et~al.}]{schnabel2001generic}
Schnabel, J.~A., Rueckert, D., Quist, M., Blackall, J.~M., Castellano-Smith,
  A.~D., Hartkens, T., Penney, G.~P., Hall, W.~A., Liu, H., Truwit, C.~L.,
  et~al., 2001. A generic framework for non-rigid registration based on
  non-uniform multi-level free-form deformations. In: Medical Image Computing
  and Computer-Assisted Intervention--MICCAI 2001. Springer, pp. 573--581.

\bibitem[{Simpson et~al.(2015)Simpson, Cardoso, Modat, Cash, Woolrich,
  Andersson, Schnabel, Ourselin, et~al.}]{simpson2015probabilistic}
Simpson, I.~J., Cardoso, M.~J., Modat, M., Cash, D.~M., Woolrich, M.~W.,
  Andersson, J.~L., Schnabel, J.~A., Ourselin, S., et~al., 2015. Probabilistic
  non-linear registration with spatially adaptive regularisation. Medical Image
  Analysis 26~(1), 203--216.

\bibitem[{Smit et~al.(2017)Smit, Lawonn, Kraima, DeRuiter, Sokooti, Bruckner,
  Eisemann, and Vilanova}]{smit2017pelvis}
Smit, N., Lawonn, K., Kraima, A., DeRuiter, M., Sokooti, H., Bruckner, S.,
  Eisemann, E., Vilanova, A., 2017. Pelvis: Atlas-based surgical planning for
  oncological pelvic surgery. IEEE Transactions on Visualization and Computer
  Graphics 23~(1), 741--750.

\bibitem[{Sokooti et~al.(2017)Sokooti, de~Vos, Berendsen, Lelieveldt,
  I{\v{s}}gum, and Staring}]{sokooti2017nonrigid}
Sokooti, H., de~Vos, B., Berendsen, F., Lelieveldt, B.~P., I{\v{s}}gum, I.,
  Staring, M., 2017. Nonrigid image registration using multi-scale 3d
  convolutional neural networks. In: International Conference on Medical Image
  Computing and Computer-Assisted Intervention. Springer, pp. 232--239.

\bibitem[{Sokooti et~al.(2016)Sokooti, Saygili, Glocker, Lelieveldt, and
  Staring}]{sokooti2016accuracy}
Sokooti, H., Saygili, G., Glocker, B., Lelieveldt, B.~P., Staring, M., 2016.
  Accuracy estimation for medical image registration using regression forests.
  In: International Conference on Medical Image Computing and Computer-Assisted
  Intervention. Springer, pp. 107--115.

\bibitem[{Staring et~al.(2014)Staring, Bakker, Stolk, Shamonin, Reiber, and
  Stoel}]{staring2014towards}
Staring, M., Bakker, M., Stolk, J., Shamonin, D., Reiber, J., Stoel, B., 2014.
  Towards local progression estimation of pulmonary emphysema using {CT}.
  Medical Physics 41~(2), 021905.

\bibitem[{Staring et~al.(2010)Staring, Klein, Reiber, Niessen, and
  Stoel}]{staring2010pulmonary}
Staring, M., Klein, S., Reiber, J.~H., Niessen, W.~J., Stoel, B.~C., 2010.
  Pulmonary image registration with elastix using a standard intensity-based
  algorithm. Medical Image Analysis for the Clinic: A Grand Challenge, 73--79.

\bibitem[{Stolk et~al.(2007)Stolk, Putter, Bakker, Shaker, Parr, Piitulainen,
  Russi, Grebski, Dirksen, Stockley, Reiber, and Stoel}]{stolk2007progression}
Stolk, J., Putter, H., Bakker, E.~M., Shaker, S.~B., Parr, D.~G., Piitulainen,
  E., Russi, E.~W., Grebski, E., Dirksen, A., Stockley, R.~A., Reiber, J.
  H.~C., Stoel, B.~C., 2007. Progression parameters for emphysema: a clinical
  investigation. Respiratory Medicine 101~(9), 1924--1930.

\bibitem[{Th{\"o}rnqvist et~al.(2010)Th{\"o}rnqvist, Petersen, H{\o}yer,
  Bentzen, and Muren}]{thornqvist2010propagation}
Th{\"o}rnqvist, S., Petersen, J.~B., H{\o}yer, M., Bentzen, L.~N., Muren,
  L.~P., 2010. Propagation of target and organ at risk contours in radiotherapy
  of prostate cancer using deformable image registration. Acta Oncologica
  49~(7), 1023--1032.

\bibitem[{Tilly et~al.(2013)Tilly, Tilly, and Ahnesj{\"o}}]{tilly2013dose}
Tilly, D., Tilly, N., Ahnesj{\"o}, A., 2013. Dose mapping sensitivity to
  deformable registration uncertainties in fractionated radiotherapy--applied
  to prostate proton treatments. BMC Medical Physics 13~(1), 2.

\bibitem[{Tustison et~al.(2013)Tustison, Song, Gee, and Avants}]{tustisontwo}
Tustison, N., Song, G., Gee, J., Avants, B., 2013. Two greedy syn variants for
  pulmonary image registration.

\bibitem[{Veiga et~al.(2015)Veiga, Louren{\c{c}}o, Mouinuddin, van Herk, Modat,
  Ourselin, Royle, and McClelland}]{veiga2015toward}
Veiga, C., Louren{\c{c}}o, A.~M., Mouinuddin, S., van Herk, M., Modat, M.,
  Ourselin, S., Royle, G., McClelland, J.~R., 2015. Toward adaptive
  radiotherapy for head and neck patients: Uncertainties in dose warping due to
  the choice of deformable registration algorithm. Medical Physics 42~(2),
  760--769.

\bibitem[{Viola and Jones(2004)}]{viola2004robust}
Viola, P., Jones, M.~J., 2004. Robust real-time face detection. International
  journal of computer vision 57~(2), 137--154.

\end{thebibliography}

\end{document}